\documentclass[12pt, draftclsnofoot, onecolumn]{IEEEtran}
\ifCLASSINFOpdf
\else
\usepackage[dvips]{graphicx}
\fi
\usepackage[cmex10]{amsmath}
\usepackage{upgreek}
\usepackage{booktabs}
\usepackage{epsfig}
\usepackage{latexsym}
\usepackage{multirow}
\usepackage{stfloats}
\usepackage{epstopdf}
\usepackage{color}  
\usepackage{tabularx} 
\usepackage{amssymb}
\usepackage{enumerate}
\graphicspath{{./Figures/}}
\usepackage{bbm}
\usepackage{bm}
\usepackage{cite}
\usepackage[tight,footnotesize]{subfigure}
\usepackage{balance}
\usepackage{mathrsfs}
\usepackage{verbatim}
\usepackage{dsfont}
\usepackage{verbatim}
\usepackage{tikz}
\usepackage{diagbox}
\usepackage{caption}
\usepackage[framemethod=tikz]{mdframed}
\usepackage{multicol}
\usepackage{environ}
\usepackage{tikz}

\begin{document}
\title{
Energy-efficient Dynamic-subarray with Fixed True-time-delay Design for Terahertz Wideband Hybrid Beamforming
}
\author{	
	Longfei Yan,
	Chong Han,~\IEEEmembership{Member,~IEEE}, and Jinhong Yuan,~\IEEEmembership{Fellow,~IEEE}
	\thanks{
	This work was accepted to present in part at the IEEE Globecom 2021~\cite{DS_FTTD_GC2021}.
				
	Longfei Yan and Chong Han are with the Terahertz Wireless Communications (TWC) Laboratory, Shanghai Jiao Tong University, Shanghai 200240, China (e-mail:\{longfei.yan, chong.han\}@sjtu.edu.cn).
	
	Jinhong Yuan is with the School of Electrical Engineering and Telecommunications, University of New South Wales, Sydney, NSW 2052, Australia (e-mail: j.yuan@unsw.edu.au).
}}
\maketitle
\markboth{}{}
\thispagestyle{empty} % no page number for the first page
\begin{abstract}
\boldmath
Hybrid beamforming for Terahertz (THz) ultra-massive multiple-input multiple-output (UM-MIMO) systems is a promising technology for 6G space-air-ground integrated networks, which can overcome huge propagation loss and offer unprecedented data rates. With ultra-wide bandwidth and ultra-large-scale antennas array in THz band, the beam squint becomes one of the critical problems which could reduce the array gain and degrade the data rate substantially.
However, the traditional phase-shifters-based hybrid beamforming architectures cannot tackle this issue due to the frequency-flat property of the phase shifters.
In this paper, to combat the beam squint while keeping high energy efficiency, a novel dynamic-subarray with fixed true-time-delay (DS-FTTD) architecture is proposed. Compared to the existing studies which use the complicated adjustable TTDs, the DS-FTTD architecture has lower power consumption and hardware complexity, thanks to the low-cost FTTDs. Furthermore, a low-complexity row-decomposition (RD) algorithm is proposed to design hybrid beamforming matrices for the DS-FTTD architecture. Extensive simulation results show that, by using the RD algorithm, the DS-FTTD architecture achieves near-optimal array gain and significantly higher energy efficiency than the existing architectures. Moreover, the spectral efficiency of DS-FTTD architecture with the RD algorithm is robust to the imperfect channel state information.
\end{abstract}
 \begin{IEEEkeywords}
 Terahertz (THz) band, hybrid beamforming, dynamic-subarray, fixed true-time-delay, beam squint.
 \end{IEEEkeywords}
\IEEEpeerreviewmaketitle
\section{Introduction}
\label{section_intro}
\subsection{Background and Motivation}
\label{section_intro_background}
As an integration of satellite systems, aerial networks, and terrestrial communications, the space-air-ground integrated network (SAGIN), has been envisioned as a critical technology in the sixth-generation (6G) systems~\cite{8368236,8760401,9482439}. In order to meet the rapid growth of wireless data rates for SAGIN in 6G, the Terahertz (THz) band with ultra-broad bandwidth has drawn much attention\cite{Gaozhen2021THzJSAC,9509294}.
However, the unprecedented multi-GHz bandwidth in the THz band comes with a cost of huge propagation loss, which drastically limits the communication distance for THz communications~\cite{8387211}. Fortunately, the sub-millimeter wavelength allows the design of array consisting of 512 and even 1024 antennas at transceivers, to enable THz ultra-massive multiple-input multiple-output (UM-MIMO) systems~\cite{AKYILDIZ201646,8765243}.
This can provide a high array gain to compensate the path loss and deal with the distance problem. Meanwhile, multiple data streams can be supported to offer a multiplexing gain and further improve the spectral efficiency of the THz systems. 
In the THz band, many hardware constraints preclude from using conventional digital beamforming, which, instead, motivates the appealing hybrid beamforming technology~\cite{THz_HBF_2021,7786122,1}.
The hybrid beamforming divides the signal processing into the digital baseband domain and analog RF domain, which can achieve high spectral efficiency while maintaining a reasonably low hardware complexity~\cite{7445130,7336494}.

Most of the existing hybrid beamforming studies consider the use of phase shifters to generate the beamforming weights~\cite{7397861,9107073,DAoSA_JSAC_2020,7914742,9110865,9139316,BSC_2018,8793242,7880698}.
To steer a beam to one target direction, the required beamforming vectors of different frequencies are frequency-proportional. However, the phase shifter can only generate the same phase for different frequencies~\cite{BSC_2018,9134775}. Using the same beamforming vector generated by phase shifters to steer the beam over a wide bandwidth, the resulted beams of different frequencies are dispersed from the target direction, which is the so-called beam squint effect~\cite{9399122,8254862,BSC_2018,8443598}.
The beam squint is a joint spatial-frequency effect~\cite{8443598}. On one hand, with larger fractional bandwidth, i.e., ratio between the bandwidth and central frequency, the beam misalignment from the target direction becomes larger. On the other hand, the larger array aperture brings a narrower beam which is more sensitive to the beam misalignment and results in higher array gain loss. 

In THz UM-MIMO wideband hybrid beamforming, due to the ultra-large fractional bandwidth and ultra-large-scale antennas array, e.g., 50 GHz bandwidth at 300 GHz with 1024 antennas, the beam squint effect is severe and the resulted array gain loss could exceed 10 dB~\cite{9398864,daiDPP}, which motivates the novel hybrid beamforming designs to combat the beam squint. 
Moreover, with a high operating frequency and a large quantity of power-hungry RF devices in THz UM-MIMO systems, the THz hybrid beamforming architecture needs to be carefully designed to reduce the power consumption and improve the energy efficiency.
Therefore, to combat the beam squint while keeping high energy efficiency, we propose a novel dynamic-subarray with fixed true-time-delay (DS-FTTD) architecture in this work.

\subsection{Related Work}
\label{section_intro_related_works}
\subsubsection{Solving beam squint in phase shifters-based architectures}
There have been some studies investigating wideband hybrid beamforming algorithms to overcome the beam squint problem for the phase shifters-based architectures, e.g., fully-connected (FC) and array-of-subarrays (AoSA) hybrid beamforming architectures.
The authors in~\cite{BSC_2018} proposed a beam squint compensation algorithm which can compensate the beamforming weights error of multiple carriers for the FC architecture. 
A wideband alternating minimization algorithm using phase extraction (PE-AltMin) was proposed to optimize the beamforming weights by jointly considering all carriers for the FC architecture~\cite{7397861}. Moreover, the authors in~\cite{8793242} proposed to utilize the knowledge of the dominant subspace of the channel matrix to design hybrid beamforming matrices to combat the beam quint. 
Apart from them, the authors in~\cite{7880698} proposed wideband hybrid beamforming algorithms which utilize the sum channel covariance matrix of all carriers for both FC and AoSA architectures.
The main principle of the above wideband algorithms is designing the digital precoding/combining matrix to re-focus the dispersed beams of all frequencies caused by the analog phase shifters.
For the microwave and mmWave systems with relatively small fractional bandwidth, the required beamforming vectors of different frequencies are similar. Hence, the beam squint is not severe and can be overcome by the above algorithms.
However, for THz communications with ultra-large fractional bandwidth and ultra-large-scale antennas, on one hand, the required beamforming vectors among the frequencies in a wide bandwidth are more distinctive such that the beam dispersion is severer, which is difficult to be re-focused by the above algorithms.
On the other hand, the beams generated by ultra-large-scale antennas array in THz systems are much narrower and more sensitive to the beam dispersion, which results in a severer array gain reduction when the beams can not be re-focused. Consequently, due to these reasons, the beam squint problem in THz UM-MIMO systems is difficult to be addressed by the above algorithms.
Instead, except for the wideband beamforming algorithm design, novel hardware design is urgently needed and motivated.

\begin{figure*}
	\centering
	\includegraphics[scale=0.45]{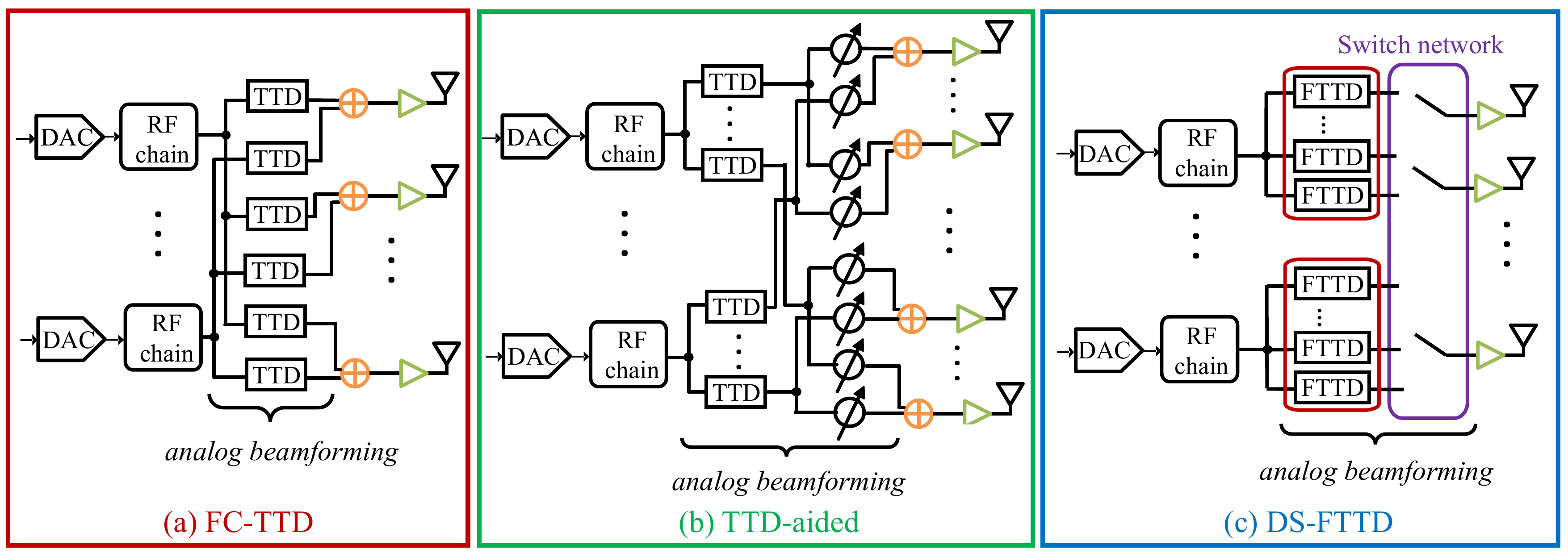} 
	\captionsetup{font={footnotesize}}
	\caption{The architectures of wideband THz hybrid beamforming (the digital part is not drawn) (a) FC-TTD, (b) TTD-aided, (c) The proposed DS-FTTD.}  
	\label{architecture_all} 
	\vspace{-9.5mm}
\end{figure*}
\subsubsection{Using true-time-delay (TTD) to solve beam squint} Different from the phase shifter, TTD can realize frequency-proportional phases, by generating the same time delay on different carriers~\cite{7959180,Gaozhen2021THzJSAC}. Utilizing the frequency-proportional property of TTD to generate the beamforming weights, the beam squint problem can be solved.
Based on this, one straightforward solution to combat beam squint is using TTDs to substitute the phase shifters~\cite{6531062,9134775,daiDPP}, e.g., the fully-connected architecture with TTD (FC-TTD) as shown in Fig.~\ref{architecture_all}(a). However, both the power consumption and hardware complexity of the TTD are higher than the phase shifter such that directly substituting all phase shifters by TTDs is impractical at THz band~\cite{9398864,daiDPP}. Instead, the authors in~\cite{9398864,daiDPP,9399122} proposed to insert a limited number of TTDs, e.g., several tens of TTDs, between the RF chains and phase shifters, to form a TTD-aided architecture, as shown in Fig.~\ref{architecture_all}(b). Due to the less number of TTDs, the TTD-aided architecture has lower power consumption than FC-TTD. However, all the above studies consider to use the TTDs which can provide adjustable time delay with high-resolution or even infinite-resolution, which are power-hungry and impractical. In light of this, we propose a novel energy-efficient DS-FTTD architecture, by using low-cost FTTD which only provides a fixed and non-adjustable time delay, to combat the beam squint. Furthermore, a switch network is proposed to enable the dynamic connections between FTTDs and antennas to enhance the spectral efficiency.

\subsection{Contributions}
In this work, we propose an energy-efficient DS-FTTD architecture by using low-cost FTTDs and a switch network, as shown in Fig.~\ref{architecture_all}(c). The concept of FTTD is similar to adjustable TTD except that its time delay is fixed rather than adjustable. Specifically, the FTTD can be realized by a microstrip line or waveguide with fixed length, which are basic devices in circuit design and have very low hardware complexity and power consumption~\cite{5481992}. However, due to the fixed time delay, the FTTD can not generate flexible beamforming weights to meet the requirement of mobile communications. To tackle this problem, we design a switch network between FTTDs and antennas to enable the dynamic connections, by which each antenna can select one FTTD with proper time delay from all FTTDs to generate flexible beamforming weights and enhance the spectral efficiency.
Furthermore, to design the hybrid beamforming matrices for the DS-FTTD architecture, we propose a low-complexity row-decomposition (RD) algorithm. 
Compared to our prior and shorter version of this work~\cite{DS_FTTD_GC2021}, more meticulous algorithm design, computational complexity analysis, and convergence analysis are proposed, with substantially more simulation evaluation.
Moreover, we add comprehensive comparisons of the power consumption and hardware complexity between the DS-FTTD architecture and the existing architectures. 
Distinctive contributions of this work are summarized as follows.
\begin{itemize}
	\item
	\textbf{We analyze the severe beam squint caused by THz peculiarities and propose an energy-efficient DS-FTTD architecture to combat beam squint.} We first analyze the severe beam squint caused by the ultra-wide fractional bandwidth and ultra-large-scale antennas array of the wideband THz UM-MIMO hybrid beamforming. Then, to combat the beam squint while maintaining a low power consumption, we propose the DS-FTTD architecture using the low-cost FTTDs and switch network. Moreover, we analyze that the power consumption, hardware complexity, and insertion loss of the proposed DS-FTTD architecture are lower than the existing FC-TTD and TTD-aided architectures.
\end{itemize} 
\begin{itemize}
	\item
	\textbf{We formulate the hybrid beamforming problem of the DS-FTTD architecture and propose a low-complexity RD algorithm to solve the hybrid beamforming matrices.} With determined time delays of FTTDs, the analog beamforming is realized by the dynamic selections enabled by the switch network matrix. The RD algorithm alternatively designs the switch network matrix and the digital beamforming matrix. Particularly, the key idea of the RD algorithm is transforming the original non-convex and intractable hybrid integer programming about the switch network matrix design to a tractable ranking problem, which can be efficiently solved. Moreover, the computational complexity of the RD algorithm is linearly related with the number of antennas, which is low.
	\end{itemize} 
\begin{itemize}
	\item
	\textbf{We carry out extensive simulations to evaluate the performance of the proposed DS-FTTD architecture and the RD algorithm.} Based on the RD algorithm, the DS-FTTD architecture achieves near-optimal array gain over all carriers. Moreover, we analyze the energy efficiency of the DS-FTTD architecture with varying transmit power and number of antennas and show that the DS-FTTD architecture achieves substantially higher energy efficiency than the existing architectures. Furthermore, the simulation results reveal that the spectral efficiency of the proposed DS-FTTD architecture with RD algorithm is robust to the imperfect channel state information (CSI).  
\end{itemize} 

The remainder of this paper is organized as follows. We investigate the beam squint effect and analyze the severe beam squint caused by THz peculiarities in Sec.~\ref{section_beam_squint}. We propose the DS-FTTD architecture and analyze its low hardware complexity and low power consumption in Sec.~\ref{section_architecture_DS_FTTD}. We formulate the hybrid beamforming problem and propose the RD algorithm for DS-FTTD in Sec.~\ref{section_RD_algorithm}. Furthermore, extensive simulation results are presented in Sec.~\ref{section_simulation}. Finally, the conclusion is drawn in Sec.~\ref{section_conclusion}.

\textbf{Notations}: 
We use upper bold case letters for matrices, lower bold case letters for vectors, and lower normal case letters for scalars;
$\textbf{I}_{N}$ denotes an $N$-dimensional identity matrix; $(\cdot)^T$, $(\cdot)^*$, and $(\cdot)^{H}$ represent transpose, conjugate, and conjugate transpose; $\lVert\cdot\rVert_{p}$ is the $p$-norm of the vector; $\lVert\cdot\rVert_{F}$ is the Frobenius norm of the matrix; $\otimes$ denotes the Kronecker product; ${\rm Re}(\cdot)$ denotes the real part; ${\rm Tr}(\cdot)$ represents the trace of the matrix; ${\rm diag}(\cdot)$ denotes the diagonal vector of the matrix.

\section{THz Wideband UM-MIMO Channel and Severe Beam Squint Caused by THz Peculiarities}
\label{section_beam_squint}
In this section, we first introduce the channel model of THz wideband UM-MIMO systems. Then, we investigate the beam squint effect and the resulted beam misalignment as well as array gain loss. More importantly, we analyze the severe beam squint in THz wideband UM-MIMO systems caused by the THz peculiarities of ultra-broad fractional bandwidth and the ultra-large-scale antennas array.
\subsection{THz Wideband UM-MIMO Channel Model}
We consider a multi-carrier wideband system with $M$ carriers, where the index $m$ represents the $m^{\rm th}$ carrier with $m=1,2,...,M$. The frequency $f_m$ of the $m^{\rm th}$ carrier is $f_m=f_c+\frac{B}{M-1}(m-\frac{M+1}{2})$, where $f_c$ and $B$ are the central frequency and the bandwidth.
We adopt $N_t$ transmitted antennas and $N_r$ received antennas. 
Multiple beams can be generated to transmit/receive the signals along the directions of the multipath components. As a result, it is necessary to consider the multipath effect.
Therefore, the channel matrix of the $m^{\rm th}$ carrier $\textbf{H}[m]\in\mathbb{C}^{N_r\times N_t}$ can be written as~\cite{DAoSA_JSAC_2020}
\begin{equation}
\textbf{H}[m]\!=\sum\nolimits_{n=1}^{N_p}\!\alpha_{n}[m]G_tG_r
\textbf{a}_{rn}[m]\textbf{a}_{tn}[m]^{H}\!\!\!,
\label{channel_model_planar}
\end{equation}%
where $N_p$ is the number of multipath of the channel and the subscript $n$ denotes the $n^{\rm th}$ multipath component. 
The complex number $\alpha_{n}[m]$ describes the path gain of the $ n^{th} $ path at the $m^{\rm th}$ carrier, which contains the free-space loss, reflection loss, scattering loss, diffraction loss, and molecular absorption loss that are detailed in our previous THz channel work~\cite{6998944}.
The vectors $ \textbf{a}_{tn}[m]$ and $ \textbf{a}_{rn}[m]$ denote the transmitted and received array response vectors for the $n^{\rm th}$ path at the $m^{\rm th}$ carrier, respectively. $\textbf{a}_{tn}[m]$ and $\textbf{a}_{rn}[m]$ share the similar structure with $\textbf{a}[m]$ as follows.
For an $L\times W$-element uniform planar array (UPA) on the $yz$-plane, $\textbf{a}[m]$ can be written as~\cite{7845674}
\begin{subequations}
\begin{align}
\textbf{a}[m]&=\big[1,..., e^{j\frac{2\pi f_m}{c}d(a {\rm sin}(\phi){\rm sin}(\theta)+b{\rm cos}(\theta))},
... ,e^{j\frac{2\pi f_m}{c}d((L-1){\rm sin}(\phi){\rm sin}(\theta)+(W-1){\rm cos}(\theta))}\big]^{\!T}\!\!\!\label{steering_vector_UPA_1_1}\\
&=\big[1,
... ,e^{j\frac{2\pi f_m}{c}d(L-1){\rm sin}(\phi){\rm sin}(\theta)}\big]^T\otimes\big[1,
... ,e^{j\frac{2\pi f_m}{c}d(W-1){\rm cos}(\theta)}\big]^T,
\end{align}
\label{steering_vector_UPA_1}%
\end{subequations}
where $d$ is the antenna spacing and $c$ denotes the speed of light.
$0\leq a\leq (L-1)$ and $0\leq b \leq (W-1)$. In addition, $\phi$ and $\theta$ refer to the azimuth and elevation angles. For $\textbf{a}_{tn}[m]$, $\phi$ and $\theta$ should be substituted by $\phi_{tn}$ and $\theta_{tn}$, i.e., the azimuth and elevation angles of departure, respectively.
Similarly, for $\textbf{a}_{rn}[m]$, $\phi$ and $\theta$ should be substituted by $\phi_{rn}$ and $\theta_{rn}$, i.e., the azimuth and elevation angles of arrival, respectively.
$G_t$ and $G_r$ represent the antenna gain at the transmitter and receiver, respectively. We adopt a widely-used sector antenna model whose antenna gain can be expressed as~\cite{9514889}
\begin{equation}
G_t=\left\{\begin{array}{ll}
\sqrt{G_{0}}, & \forall \phi_{tn} \in\left[\phi_{\min }, \phi_{\max }\right], \forall \theta_{tn} \in\left[\theta_{\min }, \theta_{\max }\right] \\
0, & \text { otherwise, }
\end{array}\right.
\label{antenna_gain_model}
\end{equation}
where $\phi_{\max }-\phi_{\min }$ and $\theta_{\max }-\theta_{\min }$ denote the antenna beamwidth at the azimuth and elevation directions, respectively, and $G_0\approx\frac{4\pi}{(\phi_{\max }-\phi_{\min })(\theta_{\max }-\theta_{\min })}$~\cite{9514889}. The antenna model at the receiver is similar to the antenna model at transmitter, with the same antenna beamwidth and gain $G_0$. 
Note that the beam of the whole antennas array is contributed by two parts, i.e., the array beam pattern and the antenna beam pattern. On one hand, the antenna beam pattern is usually fixed due to the hardware limitation. On the other hand, the array beam pattern is adjustable, thanks to the beamforming technology. The beamwidth of the whole antennas array is much narrower than the antenna beamwidth due to the beamforming enabled by the ultra-large-scale array.
Moreover, $G_0$ is the gain of the individual antenna, besides which the UM-MIMO system also has a huge array gain generated by beamforming, which is analyzed as follows.

\subsection{Beam Squint Effect}
Most of the existing hybrid beamforming studies consider the use of phase shifters to generate the beamforming weights~\cite{1,7880698,7397861,DAoSA_JSAC_2020,7914742,9110865,9139316,BSC_2018,7604098,9110889,8984302}, which suffer from the beam squint problem as follows.
We analyze the case of single-beam for an $L\times W$-element UPA, while multi-beam cases are similar and extensible.
The azimuth and elevation angle of the target beam is $\phi_0$ and $\theta_0$.
Based on~\eqref{steering_vector_UPA_1}, the array response vector for azimuth angle $\phi_0$ and elevation angle $\theta_0$ at frequency $f_m$ is
\begin{equation}
\begin{aligned}
\textbf{a}_{\rm UPA}[m]=\big[1,
... ,e^{j\frac{2\pi f_m}{c}d(L-1){\rm sin}(\phi_0){\rm sin}(\theta_0)}\big]^T\otimes\big[1,
... ,e^{j\frac{2\pi f_m}{c}d(W-1){\rm cos}(\theta_0)}\big]^T,
\end{aligned}
\label{array_response_fm_UPA}
\end{equation}
To steer a beam to $\phi_0$ and $\theta_0$, the beamforming vector composed by the weights of phase shifters is usually designed for central frequency $f_c$ as
\begin{subequations}
\begin{align}
\textbf{w}_{\rm UPA}&=\big[1,
... ,e^{j\frac{2\pi f_c}{c}d(L-1){\rm sin}(\phi_0){\rm sin}(\theta_0)}\big]^T\otimes\big[1,
... ,e^{j\frac{2\pi f_c}{c}d(W-1){\rm cos}(\theta_0)}\big]^T
\label{steering_UPA_PS_1}\\
&=\big[1,
... ,e^{j\frac{2\pi f_m}{c}d(L-1)\frac{f_c}{f_m}{\rm sin}(\phi_0){\rm sin}(\theta_0)}\big]^T\otimes\big[1,
... ,e^{j\frac{2\pi f_m}{c}d(W-1)\frac{f_c}{f_m}{\rm cos}(\theta_0)}\big]^T.
\label{steering_UPA_PS_2}
\end{align}
\label{steering_UPA_PS}%
\end{subequations}
Using this beamforming vector, the beam of central frequency $f_c$ can be steered to the target direction $\phi_0$ and $\theta_0$. However, for the other frequencies $f_m\neq f_c$, by comparing~\eqref{steering_UPA_PS_2} with the array response vector~\eqref{array_response_fm_UPA}, the actual steered spatial azimuth direction and spatial elevation direction of frequency $f_m$ are $\frac{f_c}{f_m}{\rm sin}(\phi_0){\rm sin}(\theta_0)$ and $\frac{f_c}{f_m}{\rm cos}(\theta_0)$, which reveal that the actual elevation and azimuth angles of the steered beam at frequency $f_m$ are misaligned to~\cite{9130760} 
\begin{subequations}
	\begin{align}
	&\widetilde{\theta}_0={\rm arccos}\bigg(\frac{f_c}{f_m}{\rm cos}(\theta_0)\bigg),\\
	&\widetilde{\phi}_0={\rm arcsin}\bigg(\frac{f_c{\rm sin}(\phi_0){\rm sin}(\theta_0)}{f_m{\rm sin}(\widetilde{\theta}_0)}\bigg).
	\end{align}
	\label{angle_beam_squint_UPA}%
\end{subequations}
Furthermore, by calculating the square of the inner product of the beamforming vector in~\eqref{steering_UPA_PS_1} and the array response vector in~\eqref{array_response_fm_UPA}, the array gain for frequency $f_m$ at the target direction can be calculated as~\cite{9398864,8254862,9399122}
\begin{subequations}
	\begin{align}
	G_{\rm array}[f_m,\phi_0,\theta_0]
	&=\frac{1}{LW}\big\lvert\textbf{w}_{\rm UPA}^H\textbf{a}_{\rm UPA}[m]\big\rvert^2
	\label{array_gain_UPA_1}\\
	&=\frac{1}{LW}\lvert(\textbf{v}_{a}\otimes\textbf{v}_{e})^H(\textbf{b}_{a}\otimes\textbf{b}_{e})\rvert^2
	\label{array_gain_UPA_2}\\
	&=\frac{1}{LW}\lvert(\textbf{v}_{a}^H\textbf{b}_{a})\otimes(\textbf{v}_{e}^H\textbf{b}_{e})\rvert^2
	\label{array_gain_UPA_3}\\
	&=\frac{1}{LW}\left\lvert\frac{{\rm sin}(\pi d L\Psi_a)}{{\rm sin}(\pi d \Psi_a)}\frac{{\rm sin}(\pi d W\Psi_e)}{{\rm sin}(\pi d \Psi_e)}\right\rvert^2,
	\label{array_gain_UPA_4}%
	\end{align}
\end{subequations}
where in~\eqref{array_gain_UPA_2} $\textbf{v}_a=[1,...,e^{j2\pi d\frac{f_c}{c}(L-1){\rm sin}(\phi_0){\rm sin}(\theta_0)}]^T$, $\textbf{v}_e=[1,..., e^{j2\pi d\frac{f_c}{c}(W-1){\rm cos}(\theta_0)}]^T$, $\textbf{b}_a=[1,...,\\e^{j2\pi d\frac{f_m}{c}(L-1){\rm sin}(\phi_0){\rm sin}(\theta_0)}]^T$, and $\textbf{b}_e=[1,..., e^{j2\pi d\frac{f_m}{c}(W-1){\rm cos}(\theta_0)}]^T$. \eqref{array_gain_UPA_3} follows the mixed-product property of the Kronecker product.
In~\eqref{array_gain_UPA_4}, $\Psi_a\!=\!\frac{f_m-f_c}{c}{\rm sin}(\phi_0){\rm sin}(\theta_0)$ and $\Psi_e\!=\!\frac{f_m-f_c}{c}{\rm cos}(\theta_0)$. 

Specifically, the maximal array gain equals to $LW$ and is achieved when $f_m=f_c$. While for $f_m\neq f_c$, the array gain can not achieve the maximum due to the beam misalignment. We define $L_{\rm array}[f_m,\phi_0,\theta_0]=10{\rm log}_{10}(LW)-10{\rm log}_{10}(G_{\rm array}[f_m,\phi_0,\theta_0])$ as the array gain loss with respect to the frequency $f_m$ and direction $\phi_0$ and $\theta_0$, which can be expressed as
\begin{equation}
L_{\rm array}[f_m,\phi_0,\theta_0]=\!-20{\rm log}_{10}\bigg(\frac{1}{LW}\bigg\lvert\frac{{\rm sin}\left(\pi d L\Psi_a\right)}{{\rm sin}\left(\pi d \Psi_a\right)}\frac{{\rm sin}\left(\pi d W\Psi_e\right)}{{\rm sin}\left(\pi d \Psi_e\right)}\bigg\rvert\bigg).
\label{array_gain_loss_UPA}
\end{equation} 

\subsection{Severe Beam Squint Caused by THz Peculiarities}

The beam squint is a \textbf{joint spatial-frequency effect}. In the frequency domain, the beam misalignment angle of $f_m$ in~\eqref{angle_beam_squint_UPA} grows with larger frequency deviation from the central frequency $f_c$. Compared to lower frequencies, due to the ultra-wide fractional bandwidth, the beam misalignment angle in THz systems becomes larger. Considering the spatial domain, with more antennas and a larger array aperture, the beamwidth of THz UM-MIMO systems becomes narrower, which makes the beam more sensitive to the beam misalignment in~\eqref{angle_beam_squint_UPA}, leading to a higher array gain reduction. Consequently, THz UM-MIMO systems with ultra-wide bandwidth and ultra-large-scale array suffer from a much severer beam squint problem than the microwave and mmWave MIMO systems.
 
\subsubsection{\textbf{Ultra-large fractional bandwidth}}
With abundant spectrum resource, the fractional bandwidth of THz UM-MIMO systems can be very large, e.g., 16.7\% (50 GHz bandwidth at 300 GHz central frequency). By contrast, the fractional bandwidth at microwave and mmWave systems is usually limited, e.g., 0.83\% (20 MHz at 2.4 GHz) and 3.33\% (2 GHz at 60 GHz). According to~\eqref{angle_beam_squint_UPA}, for THz wideband UM-MIMO systems, the beam misalignment caused by beam squint is much larger than the microwave and mmWave MIMO systems.

\subsubsection{\textbf{Ultra-large-scale antennas array}}
With sub-millimeter wavelength at THz band, ultra-massive antennas, e.g., 1024 antennas, can be arranged at a small footprint to provide a huge array gain to overcome the propagation loss. Moreover, to offer an additional antenna gain to further improve the coverage, directional antennas need to be used. One major property of the directional antenna is that its effective area is larger than that of the 0 dBi omnidirectional antenna and the corresponding antenna spacing is larger than the typical half-wavelength. By considering the sector antenna model as analyzed in~\eqref{antenna_gain_model}, the effective area of an antenna with directional gain $G_0$ is~\cite{9514889} 
\begin{equation}
A_{e}=\frac{\lambda_c^2}{4\pi}G_0,
\end{equation}
where $\lambda_c$ denotes the wavelength of the central frequency $f_c$. To overcome the severe path loss at THz band, $G_0$ is usually around 10 dBi in the existing THz UM-MIMO studies, e.g., 8 dBi in~\cite{8733134} and 17.8 dBi in~\cite{9514889}.
Taking $G_0=10$ as a typical value,  $A_{e}\approx(0.89 \lambda_c)^2$. Note that the physical area of the antenna is larger than $A_{e}$. The antenna spacing is the distance between the centers of two adjacent antennas and should be larger than the square root of the physical area of the antenna, i.e., $0.89\lambda_c$, to avoid the overlapping of antennas. Without loss of generality, we consider that the antenna spacing as $d=\lambda_c$ in this work\footnote{With antenna spacing larger than $0.5\lambda_c$, the array beam pattern will have grating lobe, if the array is composed by omnidirectional antennas. While for the array composed by directional antennas in this work, the directivity of antennas can suppress the grating lobe~\cite{9464678}.}.
Compared to the microwave and mmWave MIMO systems, the ultra-massive antennas and wider spacing of directional antennas in THz UM-MIMO systems bring much narrower beams, which are more sensitive to the beam misalignment caused by beam squint and result in a very large array gain loss.
\begin{figure}
	\centering
	\includegraphics[scale=0.52]{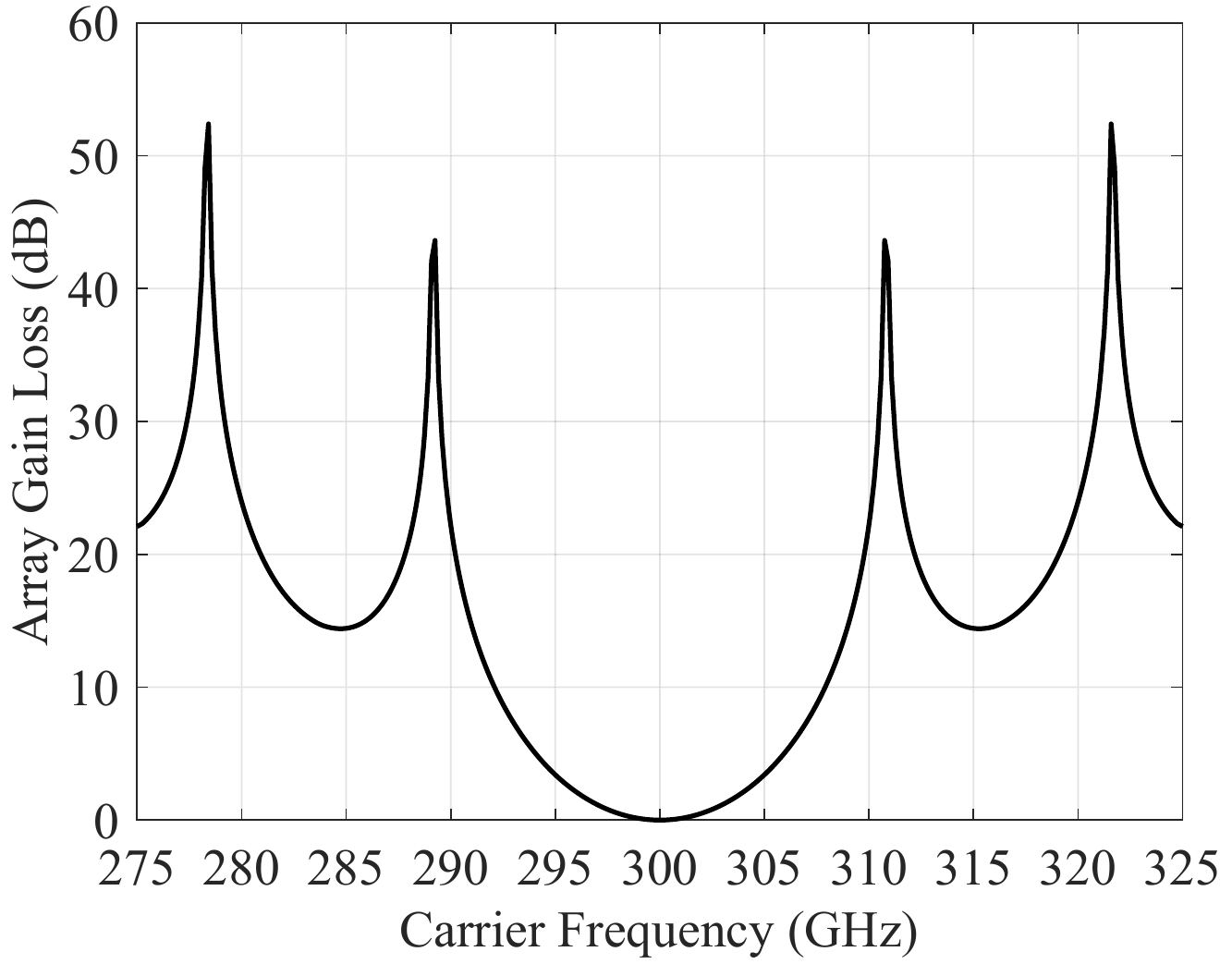} 
	\captionsetup{font={footnotesize}}
	\caption{The array gain loss caused by beam squint effect in THz UM-MIMO system with 1024 antennas. $f_c=300$ GHz, $B=50$ GHz, $L=W=32$, $\phi_0=20^{\circ}$, $\theta_0=30^{\circ}$.}  
	\label{fig_ArrayGainLoss} 
	\vspace{-6.5mm}
\end{figure}

We further numerically show the severe beam squint effect in THz wideband UM-MIMO systems in Fig.~\ref{fig_ArrayGainLoss}. There is no array gain loss at central frequency, while the array gain loss increases with $\lvert f_m-f_c\rvert$. It is worth noting that the array gain loss has a pseudo-periodicity due to the sinusoidal function in~\eqref{array_gain_loss_UPA}. In THz UM-MIMO systems, due to the ultra-wide fractional bandwidth and ultra-large-scale antennas array, the array gain loss caused by beam squint is huge, e.g., $22$ dB at $275$ GHz. Consequently, the beam squint in THz wideband UM-MIMO systems is very severe and needs to be addressed carefully.

\section{Combat the Beam Squint Energy-efficiently: The Proposed DS-FTTD Architecture}
\label{section_architecture_DS_FTTD}
To solve the severe beam squint in THz wideband systems, the existing THz hybrid beamforming studies usually consider to use the adjustable TTDs.
However, the adjustable TTD has high power consumption and hardware complexity at the THz band, which is impractical. Therefore,
to address the severe beam squint while keeping a low power consumption and hardware complexity, we propose a novel DS-FTTD hybrid beamforming architecture, using the low-cost FTTDs, in this section.

\subsection{Drawbacks of Adjustable TTD}
As analyzed in Sec.~\ref{section_beam_squint}-B, the essence of solving beam squint is providing frequency-proportional phase as the beamforming weight. Different from the frequency-flat phase shifter, the TTD adjusts the same time delay $\tau$ on the signal with different frequencies such that the adjusted phase is $2\pi f_m\tau$ and is frequency-proportional~\cite{7959180,Gaozhen2021THzJSAC}.
As we analyzed in related work, to solve the severe beam squint problem, the existing THz hybrid beamforming studies consider to substitute all phase shifters by TTDs, e.g., the FC-TTD architecture in Fig.~\ref{architecture_all}(a), or insert an additional TTD layer between RF chains and phase shifters, e.g., the TTD-aided architecture in Fig.~\ref{architecture_all}(b)~\cite{9398864,daiDPP}. However, both of these two architectures consider the use of TTD which can provide adjustable time delay with high-resolution or even infinite-resolution, which has high power consumption and hardware complexity as follows.

\begin{figure*}
	\centering
	\includegraphics[scale=0.45]{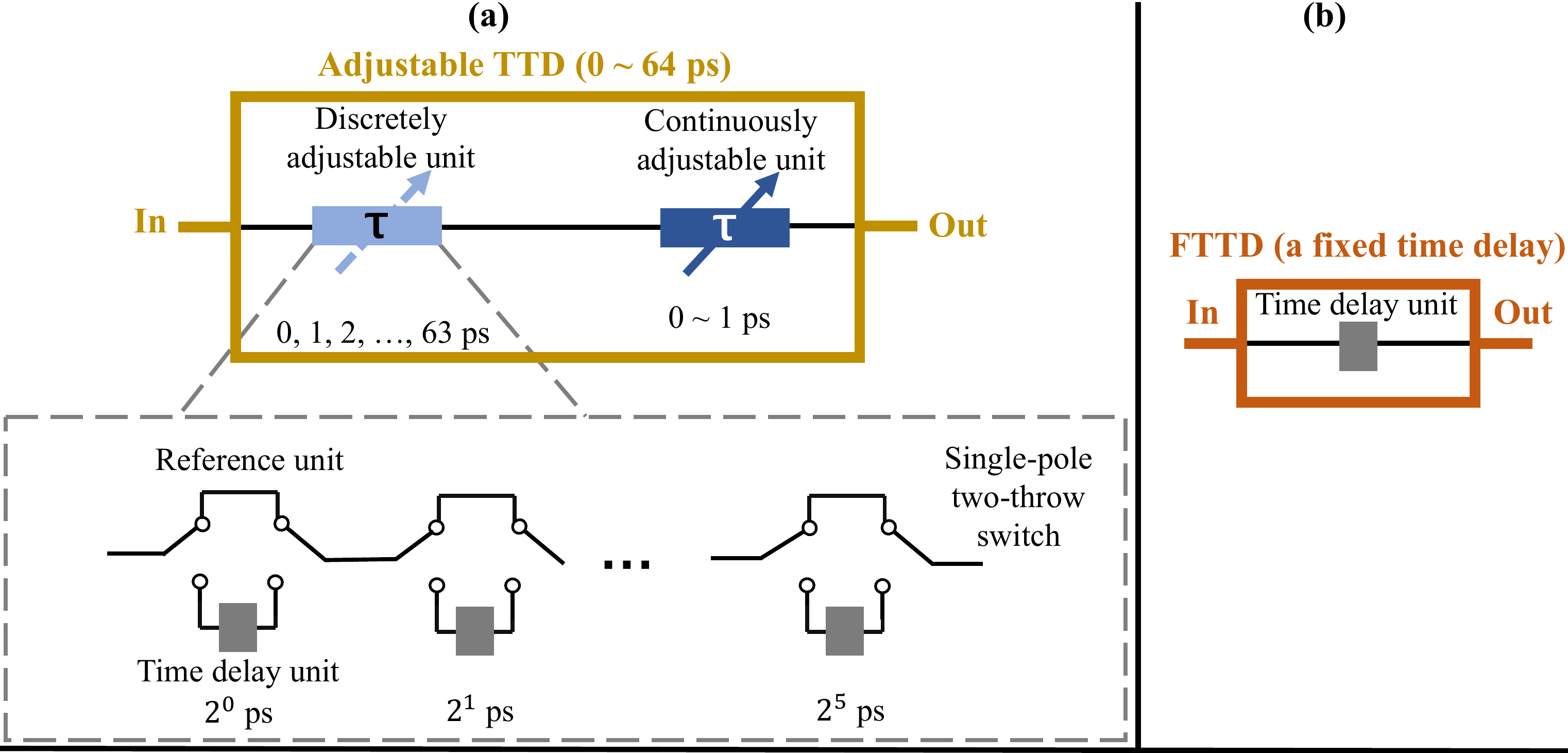} 
	\captionsetup{font={footnotesize}}
	\caption{Typical hardware implementations of adjustable TTD and FTTD, which provide (a) continuously adjustable delay in $0\sim64$ ps and (b) a fixed time delay, respectively.}  
	\label{structure_TTD} 
	\vspace{-9.5mm}
\end{figure*}
One typical hardware implementation of adjustable TTD is shown in Fig.~\ref{structure_TTD}(a)~\cite{9426936,8248806}. The TTD is composed by cascaded discretely adjustable unit and continuously adjustable unit, which can provide discrete levels of time delay in a large range and a continuously adjustable time delay in a small range, respectively. The continuously adjustable unit can be realized by movable parallel-coupled high-resistivity silicon slabs~\cite{9426936}. The discretely adjustable unit is realized by cascaded time delay units and single-pole two-throw switches, which provide different levels of time delay and time delay selection, respectively.
Due to the cascaded structure, the power consumption, insertion loss, and hardware complexity of the TTD are summation of those of the continuously adjustable unit, time delay units, and switches. As a result, the TTD at THz band usually has high power consumption, insertion loss, and hardware complexity, e.g. 80 mW power consumption and 10 dB insertion loss~\cite{Tan2020PrecodingSurvey,9398864}.

\subsection{The Proposed DS-FTTD Architecture}
To combat the beam squint problem while keeping high energy efficiency and low hardware complexity, we propose the DS-FTTD architecture, which uses simple FTTDs rather than the complicated adjustable TTDs, as shown in Fig.~\ref{architecture_all}(c). Since the FTTD provides a fixed and non-adjustable time delay, it is only composed by a simple time delay unit, as shown in Fig.~\ref{structure_TTD}(b). Specifically, the time delay unit can be realized by a microstrip line or waveguide with fixed length, which is a basic and widely-used device to convey the RF signal in the circuit~\cite{5481992}. Compared to the adjustable TTD, the power consumption and hardware complexity are substantially reduced.

In DS-FTTD architecture, each RF chain connects to $Q$ FTTDs, whose time delays are fixed as $\tau_1,\tau_2,...,\tau_Q$, respectively.
With the advantage of low cost, one drawback of the FTTD is that the provided time delay is fixed while the required beamforming weight of phase may be an arbitrary value between $0$ and $2\pi$ depending on the channel. Therefore, we propose to insert a switch network between the FTTDs and antennas to enable the dynamic connections. Through the switch, each antenna can select one FTTD with proper time delay from all FTTDs to generate the beamforming weight to satisfy the requirement of channel. Compared to the existing FC-TTD and TTD-aided architectures, the proposed DS-FTTD architecture has lower hardware complexity, power consumption, and insertion loss, as follows.

\subsubsection{\textbf{Low hardware complexity}}
The major difference among the proposed DS-FTTD and the existing FC-TTD and TTD-aided architectures is the analog beamforming part as shown in Fig.~\ref{architecture_all}, while the other parts are similar.
Therefore, we list the hardware components and the typical values of power consumption of the analog beamforming part of DS-FTTD, FC-TTD, and TTD-aided architectures in TABLE~\ref{hardware_comparison}, where $N_t$ and $L_t$ denote the number of antennas and RF chains in all architectures, $N_k$ is the number of TTDs in TTD-aided architecture, $Q$ is the number of FTTDs connected with each RF chain in DS-FTTD architecture. 
The involved hardware devices of the analog beamforming part of the FC-TTD, TTD-aided, and DS-FTTD architectures are $N_tL_t$ adjustable TTDs, $N_tL_t$ phase shifters plus $N_k$ adjustable TTDs, and $N_t$ switches plus $L_tQ$ FTTDs, respectively.	
In this work, to achieve high energy efficiency, the value of $Q$ is small, e.g., $Q=32\ll N_t$ in the simulations of the manuscript. Compared to an adjustable TTD or an adjustable phase shifter, a FTTD or a switch is of lower cost and requires less space. In terms of quantities, the combined quantity of FTTDs and switches in the DS-FTTD architecture is equal to $L_tQ+N_t$, which is much smaller than the number of adjustable TTDs $N_tL_t$ in FC-TTD architecture and the number of adjustable phase shifters $N_tL_t$ in TTD-aided architecture~\cite{daiDPP}. Consequently, without loss of generality, the hardware complexity and the required space of the DS-FTTD architecture are lower than the FC-TTD and TTD-aided counterparts. Moreover, the value of $Q$ could be further decreased to occupy less space, when applying the DS-FTTD architecture in the mobile scenarios, e.g., UAV.

\begin{table}
	\centering
	\setlength{\abovecaptionskip}{1pt}
	\captionsetup{font={footnotesize}}
	\caption{The hardware components and power consumption of the analog beamforming part of the hybrid beamforming architectures, around 300 GHz~\cite{DS_FTTD_GC2021,DAoSA_JSAC_2020,Tan2020PrecodingSurvey,6005342}.}
	\begin{tabular}{|c|cc|cc|cc|cc||p{60pt}<{\centering}|}
		\hline \multirow{2}*{}&\multicolumn{2}{c|}{Phase Shifter}&\multicolumn{2}{c|}{Adjustable TTD}&\multicolumn{2}{c|}{FTTD}&\multicolumn{2}{c||}{Switch}&Total Power\\
		\cline{2-9} &Quantity&Power&Quantity&Power&Quantity&Power&Quantity&Power&[mW]\\
		\hline \textbf{FC-TTD}&-&-&$N_tL_t$&80 mW&-&-&-&-&80$N_tL_t$\\
		\hline \textbf{TTD-aided}&$N_tL_t$&42 mW&$N_k$ &80 mW&-&-&-&-&42$N_tL_t+$80$N_k$\\
		\hline \textbf{DS-FTTD}&-&-&-&-&$L_tQ$&30 mW&$N_t$&10 mW&10$N_t+$30$\rm N^{a}_{FTTD}$\\		
		\hline
	\end{tabular}
	\label{hardware_comparison}
	\vspace{-6.5mm}
\end{table}
\subsubsection{\textbf{Low power consumption}}
We list the typical values of the power consumption of devices in TABLE~\ref{hardware_comparison}.
In DS-FTTD, FC-TTD, and TTD-aided architectures, the typical values of $N_t$, $L_t$, $N_k$, and $Q$ are 1024, 4, 128, and 32, respectively. Consequently, the power consumption of the analog beamforming part of the FC-TTD and TTD-aided architectures are $80N_tL_t=327$ W and $42N_tL_t+80N_k=182$ W, respectively. 
In the DS-FTTD architecture, although the total number of FTTDs is $L_tQ$, only the FTTDs which are selected by at least one switch are active and consume power. Hence, we denote $\rm N^{a}_{FTTD}$ as the number of active FTTDs, which is no larger than $L_tQ$. As a result, the power consumption of the analog beamforming part of the DS-FTTD architecture is $10N_t+30\rm N^{a}_{FTTD}$, as shown in TABLE~\ref{hardware_comparison}. Since we usually have $N_t\gg L_tQ\geq \rm N^{a}_{FTTD}$ in THz DS-FTTD architecture, the switches dominate the power consumption of the analog beamforming part. We have $10N_t+30{\rm N^{a}_{FTTD}}\leq 10N_t+30L_tQ=14$ W. Therefore, the power consumption of the analog beamforming part of DS-FTTD architecture is at most 4\% and 11\% of the existing FC-TTD and TTD-aided architectures. Since the other parts of these architectures are similar, the overall power consumption of the DS-FTTD is significantly lower than the FC-TTD and TTD-aided architectures.

\subsubsection{\textbf{Low insertion loss}}
The insertion loss denotes the attenuation of the signal when passing the hardware device. The transmit power of the hardware architecture in dB can be treated as the output power of signal source plus the gain of power amplifier and then minus the insertion loss. Due to the hardware challenges at THz band, the output power of signal source and the gain of power amplifier are smaller than the lower frequencies. As a result, to avoid low transmit power, the insertion loss should be reduced.
As shown in Fig.~\ref{architecture_all}, for the DS-FTTD architecture, the insertion loss of the analog beamforming part is the summation of the insertion loss of one FTTD and one switch, i.e., $IL_{\rm FTTD}+IL_{\rm switch}$. The insertion loss of the analog beamforming part of FC-TTD architecture is the insertion loss of the adjustable TTD, i.e., $IL_{\rm TTD}$. For TTD-aided architecture, the insertion loss is the summation of insertion loss of one adjustable TTD and one phase shifter, i.e., $IL_{\rm TTD}+IL_{\rm PS}$. Note that the adjustable TTD is usually cascaded by multiple FTTDs and switches, as shown in Fig.~\ref{structure_TTD}. As a result, the insertion loss of adjustable TTD is the summation of multiple FTTDs and switches such that $IL_{\rm TTD}$ is several times of $IL_{\rm FTTD}+IL_{\rm switch}$. Hence, $IL_{\rm TTD}+IL_{\rm PS}>IL_{\rm TTD}>IL_{\rm FTTD}+IL_{\rm switch}$. Accordingly, the insertion loss of the analog beamforming part of the DS-FTTD architecture is lower than those of the FC-TTD and TTD-aided architectures.

\section{System Model and Hybrid Beamforming Design for DS-FTTD Architecture}
\label{section_RD_algorithm}
In the previous section, we have proposed the energy-efficient DS-FTTD architecture and analyzed its low power consumption and hardware complexity.
In this section, we first formulate the hybrid beamforming problem for the DS-FTTD architecture. Then, we propose a low-complexity RD algorithm to solve the hybrid beamforming problem. The key idea of the RD algorithm is transforming the non-convex and intractable hybrid integer programming design problem to a tractable ranking problem, which can be efficiently solved.

\subsection{Problem Formulation for DS-FTTD Hybrid Beamforming}
To focus on the analysis of the DS-FTTD architecture at the transmitter, we consider that the receiver is arranged with optimal digital combining architecture, which is a common practice in the hybrid beamforming studies~\cite{7397861,DAoSA_JSAC_2020}, since the analysis of the DS-FTTD architecture at the receiver side is similar and extensible. The number of antennas at transmitter and receiver is $N_t$ and $N_r$, respectively. We denote the number of RF chains and data streams at transmitter as $L_t$ and $N_s$, respectively.
The system model of the DS-FTTD architecture at the $m^{\rm th}$ carrier can be expressed as
\begin{equation}
\textbf{y}[m] =  \textbf{H}[m]\textbf{S}\textbf{F}[m]\textbf{D}[m]\textbf{x}[m] +  \textbf{n}[m],
\label{system_model_DS}
\end{equation}
where $\textbf{x}[m]$ and $\textbf{y}[m]$ denote $N_s\times1$ and $N_r\times 1$ transmitted and received signals. Moreover, $\textbf{n}[m]\in\mathbb{C}^{N_r\times 1}$ refers to the noise vector, and $\textbf{F}[m]$ is an $L_tQ\times L_t$-dimensional matrix which represents the phase adjustments of the FTTDs. $\textbf{S}$ is an $N_t\times L_tQ$-dimensional binary switch network matrix. Since the state of switch is frequency-independent, we omit the frequency index $[m]$ for $\textbf{S}$ in the following. $\textbf{D}[m]\in\mathbb{C}^{L_t\times N_s}$ denotes the digital precoding matrix. We consider that the CSI has been obtained, while the impact of imperfect CSI is numerically evaluated in Sec.~\ref{section_simulation}.
In this work, we focus on the hybrid beamforming at transmitter, while one potential future extension is the use of reconfigurable intelligent surfaces (RIS) to manipulate the communication environment, which not only improves the coverage distance~\cite{8741198} but also reduces the complexity of beamforming at the transmitter~\cite{9410457}.

\textit{\textbf{Time delays of FTTDs}:} Each RF chain connects to $Q$ FTTDs, whose time delays are set as $\tau_1,\tau_2, ..., \tau_Q$, respectively. We use $\tau_{\rm min}$ and $\tau_{\rm max}$ to denote the minimal and maximal required time delays of all antennas in the array. As shown in~\eqref{steering_vector_UPA_1}, when steering a beam to the azimuth angle $\phi$ and elevation angle $\theta$, $\tau_{\rm min}=0$ and $\tau_{\rm max}=\frac{d}{c}((L-1){\rm sin}(\phi){\rm sin}(\theta)+(W-1){\rm cos}(\theta))$. With different beam directions, $\tau_{\rm max}$ varies while it is always no larger than $\frac{d}{\sqrt{2}c}(L+W-2)$ due to the property of sine function and cosine function. Hence, we set $\tau_{\rm max}=\frac{d}{\sqrt{2}c}(L+W-2)$. In mobile communications with arbitrary targeted beam directions, the required time delay of one antenna in the array can be arbitrary value between $\tau_{\rm min}$ and $\tau_{\rm max}$. Note that each antenna selects one FTTD among all FTTDs through switch to satisfy its required time delay. Therefore, to satisfy the required time delays of all antennas as far as possible, we uniformly set the time delays of FTTDs as $\tau_q=\tau_{\rm min}+\frac{\tau_{\rm max}-\tau_{\rm min}}{Q-1}(q-1)$ with $q=1,2,...,Q$, without loss of generality.

The structure of the FTTD network matrix $\textbf{F}[m]$ can be stated as
\begin{equation}
\textbf{F}[m]={\rm blkdiag}(\ \! \underbrace{\textbf{f}[m],...,\textbf{f}[m]}_{L_t}\ \! ),
\label{structure_FPS_network}
\end{equation}
where $\textbf{f}[m]=[e^{j2\pi f_m\tau_1},e^{j2\pi f_m\tau_2},...,e^{j2\pi f_m\tau_Q}]^T$ and ${\rm blkdiag}(\cdot)$ denotes block diagonal matrix. Since each antenna only selects one FTTD to connect with through the switch, the switch network matrix $\textbf{S}$ has a constraint that each row of $\textbf{S}$ only has one non-zero element, i.e., $\lVert\textbf{S}_i\rVert_{0}=1, i=1,2,...,N_t$, where $\textbf{S}_{i}$ represents the $i^{\rm th}$ row of $\textbf{S}$. $\rho$ is the total transmit power of the DS-FTTD architecture.
The achievable spectral efficiency of the DS-FTTD architecture is given by
\begin{align}
SE=\frac{1}{M}\sum\nolimits_{m=1}^{M}{\rm{log}}_2\Big(\Big\lvert\textbf{I}_{N_r}+\frac{1}{\sigma^{2}_{n}}\textbf{H}[m]\textbf{S}\textbf{F}[m]\textbf{D}[m](\textbf{H}[m]\textbf{S}\textbf{F}[m]\textbf{D}[m])^H\Big\rvert\Big),
\label{SE_formulation}
\end{align}
where $\sigma^{2}_{n}$ denotes the noise power of each carrier.

\textit{\textbf{Hybrid beamforming problem}:}
The unknown variables to design in the DS-FTTD architecture include the switch network matrix $\textbf{S}$ and the digital precoding matrix $\textbf{D}[m]$.
We aim to design $\textbf{S}$ and $\textbf{D}[m]$ to maximize the achievable spectral efficiency \eqref{SE_formulation} of the DS-FTTD architecture. Note that the energy efficiency is the ratio between spectral efficiency and power consumption.
As shown in TABLE~\ref{table_power_existing_architectures} in Sec.~\ref{section_simulation}, the power consumption of DS-FTTD architecture is dominated by ${\rm P_{PA}}N_t+{\rm P_{SW}}N_t$, which is a constant. Hence, by maximizing the spectral efficiency while keeping an almost invariant power consumption, the energy efficiency is improved.
However, directly solving $\textbf{S}$ and $\textbf{D}[m]$ to maximize the spectral efficiency is intractable, since they are coupled in \eqref{SE_formulation} and the expression of the spectral efficiency is quite complicated.
Instead, it is a common practice to transform the maximization of the achievable spectral efficiency into a Euclidean distance minimization problem~\cite{1,7397861,DAoSA_JSAC_2020}. 
Therefore, the Euclidean distance minimization problem across all carrier frequencies can be formulated as
\begin{subequations}
	\begin{align}
	&\mathop{\rm min\ }\limits_{\textbf{S},\textbf{D}[m]}\sum\nolimits_{m=1}^{M}\lVert \textbf{P}[m]-\textbf{S}\textbf{F}[m]\textbf{D}[m]\rVert_{F}^2
	\label{problem_SFD_objective}\\& 
	\mathrm{s.t.}\quad\ \textbf{S}_{i,l}\in\{0,1\}, \lVert\textbf{S}_{i}\rVert_{0}=1, \forall i,l,
	\label{problem_SFD_constraints_2}\\
	& \qquad \ \lVert\textbf{S}\textbf{F}[m]\textbf{D}[m]\rVert^{2}_{F}=\lVert\textbf{P}[m]\rVert_{F}^2, \forall m,
	\label{problem_SFD_constraints_3}
	\end{align}
	\label{problem_SFD}%
\end{subequations}
where $\textbf{P}[m]=\textbf{V}_{N_s}[m]\bm{\Gamma}[m]$ denotes the optimal precoding matrix for the $m^{\rm th}$ carrier frequency which can maximize the spectral efficiency. $\textbf{V}_{N_s}[m]$ is the first $N_s$ columns of $\textbf{V}[m]$, which comes from the singular value decomposition (SVD) of the channel matrix $\textbf{H}[m]$ such that $\textbf{H}[m]=\textbf{U}[m]\bm{\Sigma}[m]\textbf{V}[m]^H$. $\bm{\Gamma}[m]$ refers to the water-filling power allocation matrix for the $m^{\rm th}$ carrier, which satisfies $\sum_{m=1}^{M}\lVert\textbf{P}[m]\rVert_{F}^2=\rho$. $\textbf{S}_{i,l}$ is the element of $\textbf{S}$ on the $i^{\rm th}$ row and $l^{\rm th}$ column. \eqref{problem_SFD_constraints_2} and \eqref{problem_SFD_constraints_3} are the switch network constraint and transmit power constraint, respectively.
As we will see that the numerical results in Sec.~\ref{section_simulation}-D show that the transformation of the original spectral efficiency maximization to the Euclidean distance minimization \eqref{problem_SFD} is effective, i.e., by reducing the objective function in \eqref{problem_SFD_objective}, the spectral efficiency is enlarged efficiently.

\subsection{RD Hybrid Beamforming Algorithm for DS-FTTD Architecture}
We propose a low-complexity RD algorithm to solve the problem \eqref{problem_SFD}, which is a non-convex hybrid integer programming problem. 
One main difficulty to solve the problem \eqref{problem_SFD} is the coupling of $\textbf{S}$ and $\textbf{D}[m]$. 
To tackle the obstacle of the coupling, we aim to alternatively solve $\textbf{S}$ and $\textbf{D}[m]$, i.e., alternatively fix one to update the other one. Particularly, we transform the intractable hybrid integer programming problem into a tractable ranking problem, which can be efficiently solved. 
In the existing hybrid beamforming studies~\cite{1,7397861,DAoSA_JSAC_2020,9110865}, it is a common and effective practice to first design the hybrid beamforming matrices by omitting the power constraint and then normalize the digital beamforming matrix to satisfy the power constraint. Hence, in the following, we first design $\textbf{S}$ and $\textbf{D}[m]$ without the power constraint and then normalize $\textbf{D}[m]$ as $\textbf{D}[m]=\frac{\lVert\textbf{P}[m]\rVert_{F}}{\ \ \left\lVert{\textbf{S}\textbf{F}[m]\textbf{D}[m]}\right\rVert_{F}}\textbf{D}[m]$ to satisfy $\lVert\textbf{S}\textbf{F}[m]\textbf{D}[m]\rVert^{2}_{F}=\lVert\textbf{P}[m]\rVert_{F}^2, \forall m$.

\subsubsection{\textbf{Design of switch network matrix $\textbf{\rm\bf S}$}}
To begin with, we first design $\textbf{S}$ to solve problem \eqref{problem_SFD}, with fixed $\textbf{D}[m]$. By omitting the transmit power constraint temporarily, solving $\textbf{S}$ to minimize \eqref{problem_SFD_objective} is rearranged as
\begin{subequations}
\begin{align}
&\mathop{\rm min\ }\limits_{\textbf{S}}\sum\nolimits_{m=1}^{M}\lVert \textbf{P}[m]-\textbf{S}\textbf{F}[m]\textbf{D}[m]\rVert_{F}^2
\label{subproblem_SFD_objective}\\& 
\mathrm{s.t.}\quad\ \textbf{S}_{i,l}\in\{0,1\}, \
\lVert\textbf{S}_{i}\rVert_{0}=1, \forall i,l,
\label{subproblem_SFD_constraint_1}
\end{align}
\label{subproblem_SFD}%
\end{subequations}
where the problem \eqref{subproblem_SFD} is an integer programming problem associated with a matrix variable, which is inefficient to solve. To make the problem more tractable, we rewrite \eqref{subproblem_SFD_objective} as
\begin{subequations}
	\begin{align}
		&\sum\nolimits_{m=1}^{M}\left\lVert \textbf{P}[m]-\textbf{S}\textbf{F}[m]\textbf{D}[m]\right\rVert_{F}^2\\
		=&\sum\nolimits_{m=1}^{M}{\rm Tr}\left((\textbf{P}[m]-\textbf{S}\textbf{F}[m]\textbf{D}[m])(\textbf{P}[m]-\textbf{S}\textbf{F}[m]\textbf{D}[m])^H\right)\\
		=&\sum\nolimits_{m=1}^{M}\!\!{\rm Tr}\left(\textbf{P}[m]\textbf{P}[m]^H\right)\!\!-\!\!2{\rm Tr}\left({\rm Re}(\textbf{S}\textbf{F}[m]\textbf{D}[m]\textbf{P}[m]^H)\right)\!+\!{\rm Tr}\!\left(\textbf{S}\textbf{F}[m]\textbf{D}[m]\textbf{D}[m]^H\textbf{F}[m]^H\textbf{S}^H\right),
		\label{trace_expression_1}
	\end{align}
\end{subequations}
where the first term of \eqref{trace_expression_1} is a fixed value since $\textbf{P}[m]$ is known. Therefore, to minimize \eqref{subproblem_SFD_objective} is equivalent to minimize the second and third terms of \eqref{trace_expression_1}, as
\begin{equation}
\mathop{\rm min\ }\limits_{\textbf{S}}\sum\nolimits_{m=1}^{M}\Big(-\!2{\rm Tr}\left({\rm Re}(\textbf{S}\textbf{F}[m]\textbf{D}[m]\textbf{P}[m]^H)\right)\!+\!{\rm Tr}\!\left(\textbf{S}\textbf{F}[m]\textbf{D}[m]\textbf{D}[m]^H\textbf{F}[m]^H\textbf{S}^H\right)\!\Big)
\label{second_third_term}
\end{equation}
First, we rewrite $\sum\nolimits_{m=1}^{M}-2{\rm Tr}\left({\rm Re}(\textbf{S}\textbf{F}[m]\textbf{D}[m]\textbf{P}[m]^H)\right)$ as
\begin{subequations}
	\begin{align}
    &\sum\nolimits_{m=1}^{M}\!\!-2{\rm Tr}\left({\rm Re}(\textbf{S}\textbf{F}[m]\textbf{D}[m]\textbf{P}[m]^H)\right)\\
	=&-2\!\sum\nolimits_{m=1}^{M}\sum\nolimits_{i=1}^{N_t}{\rm Re}(\textbf{S}_i\textbf{F}[m]\textbf{D}[m]\textbf{P}_i[m]^H)
    \label{trace_expression_2_step2}\\
	=&-2 \!\sum\nolimits_{i=1}^{N_t} \textbf{S}_i\left(\sum\nolimits_{m=1}^{M}{\rm Re}(\textbf{F}[m]\textbf{D}[m]\textbf{P}_i[m]^H)\right),
	\label{trace_expression_2}
	\end{align}
\end{subequations}
where \eqref{trace_expression_2_step2} comes from the property of matrix trace. $\textbf{S}_i$ and $\textbf{P}_i[m]$ are the $i^{\rm th}$ row of $\textbf{S}_i$ and $\textbf{P}_i[m]$, respectively. Furthermore, we derive $\sum\nolimits_{m=1}^{M}{\rm Tr}\!\left(\textbf{S}\textbf{F}[m]\textbf{D}[m]\textbf{D}[m]^H\textbf{F}[m]^H\textbf{S}^H\right)$ in \eqref{second_third_term} as
\begin{subequations}
	\begin{align}
&\sum\nolimits_{m=1}^{M}\!{\rm Tr}\!\left(\textbf{S}\textbf{F}[m]\textbf{D}[m]\textbf{D}[m]^H\textbf{F}[m]^H\textbf{S}^H\right)
	\label{trace_expression_3_1}\\
	\!\!=\!&\sum\nolimits_{m=1}^{M}\!\sum\nolimits_{i=1}^{N_t}\!\textbf{S}_i\textbf{F}[m]\textbf{D}[m]\textbf{D}[m]^H\textbf{F}[m]^H\textbf{S}_i^H
		\label{trace_expression_3_2}\\
	\!\!=\!&\sum\nolimits_{m=1}^{M}\sum\nolimits_{i=1}^{N_t}\textbf{S}_i\ \!\!{\rm diag}\!\left(\textbf{F}[m]\textbf{D}[m]\textbf{D}[m]^H\textbf{F}[m]^H\right)\!
\label{trace_expression_3_3}\\		
	\!\!=\!&\sum\nolimits_{i=1}^{N_t}\!\!\textbf{S}_i\!\left(\sum\nolimits_{m=1}^{M}\!\!\!\!\!{\rm diag}\!\left(\textbf{F}[m]\textbf{D}[m]\textbf{D}[m]^H\textbf{F}[m]^H\right)\!\right)\!,
	\label{trace_expression_3_4}
	\end{align}
\end{subequations}
where~\eqref{trace_expression_3_2} follows the property of matrix trace. Moreover, due to the property of $\textbf{S}_i$ such that only one element of $\textbf{S}_i$ is `1' and the other elements are `0', \eqref{trace_expression_3_2} is further equivalent to~\eqref{trace_expression_3_3}.

By substituting \eqref{trace_expression_2} and \eqref{trace_expression_3_4} in \eqref{second_third_term}, solving $\textbf{S}$ to minimize the objective function \eqref{subproblem_SFD_objective} is equivalent to minimize
\begin{equation}
	\sum\nolimits_{i=1}^{N_t}\textbf{S}_i\Big(\sum\nolimits_{m=1}^{M}\Big(-2{\rm Re}(\textbf{F}[m]\textbf{D}[m]\textbf{P}_i[m]^H)+{\rm diag}\left(\textbf{F}[m]\textbf{D}[m]\textbf{D}[m]^H\textbf{F}[m]^H\right)\Big)\Big).
	\label{combination_second_and_thrid}
\end{equation} 
We point out that the design of each $\textbf{S}_i$ only influences the value of the $i^{\rm th}$ term of the summation in~\eqref{combination_second_and_thrid}, which reveals that the minimization of \eqref{combination_second_and_thrid} can be decomposed as $N_t$ uncorrelated parallel subproblems as
\begin{subequations}
	\begin{align}
	&\mathop{\rm min\ }\limits_{\textbf{S}_i}\textbf{S}_i\left(\sum\nolimits_{m=1}^{M}\left(-2{\rm Re}(\textbf{F}[m]\textbf{D}[m]\textbf{P}_i[m]^H)+{\rm diag}\left(\textbf{F}[m]\textbf{D}[m]\textbf{D}[m]^H\textbf{F}[m]^H\right)\right)\right)
	\label{subproblem_SFD_2_obj}
\\& 
	\mathrm{s.t.}\quad\ \textbf{S}_{i,l}\in\{0,1\},\ \lVert\textbf{S}_{i}\rVert_{0}=1, \forall l.
	\end{align}
	\label{subproblem_SFD_2}%
\end{subequations}
Following the binary property of the row vector $\textbf{S}_i$, i.e., $\textbf{S}_{i,l}\in\{0,1\}, \forall l$ and $ \lVert\textbf{S}_{i}\rVert_{0}=1$, minimizing \eqref{subproblem_SFD_2_obj} is equivalent to finding the position of the minimal element of the known column vector $\sum\nolimits_{m=1}^{M}\left(-2{\rm Re}(\textbf{F}[m]\textbf{D}[m]\textbf{P}_i[m]^H)+{\rm diag}\left(\textbf{F}[m]\textbf{D}[m]\textbf{D}[m]^H\textbf{F}[m]^H\right)\right)$, which is a simple ranking problem and can be solved efficiently by the sorting algorithm in the solvers. Consequently, by denoting $p_{\rm min}$ as the position of the minimal value, the optimal solution of $\textbf{S}_i$ for the problem \eqref{subproblem_SFD_2} is 
\begin{equation}
	\textbf{S}_i=[\ \underbrace{0,...,0}_{p_{\rm min}-1},\ 1,\underbrace{0,...,0}_{L_tQ-p_{\rm min}}].
	\label{solution_S_i}
\end{equation}
Then, by solving problem \eqref{subproblem_SFD_2} with $i=1,...,N_t$ in parallel, the optimal $\textbf{S}$ in problem \eqref{subproblem_SFD} is obtained. 
Since $\textbf{P}_i[m]$ with different $i$ differs from each other, the designed $\textbf{S}_i$ with different $i$ is in general different. Consequently, different antennas usually select different FTTDs to connect with. In the simulations of this work, the number of antennas is usually larger than the number of FTTDs, which leads to that some antennas may select the same FTTD with the same delay.

\subsubsection{\textbf{Design of digital precoding matrix $\textbf{\rm\bf D}[m]$}}
After determining $\textbf{S}$, we design $\textbf{D}[m]$ by fixing $\textbf{S}$ as follows.
A semi-unitary digital precoding matrix can mitigate the interference of different data streams and improve the spectral efficiency~\cite{DAoSA_JSAC_2020,7397861}.
Inspired by this, we add a semi-unitary constraint to the digital precoding matrix as $\textbf{D}[m]^{H}\textbf{D}[m]=\textbf{I}_{N_s}$ for each $m$.
By fixing $\textbf{S}$ and omitting the transmit power constraint temporarily, the design problem \eqref{problem_SFD} can be reformulated as
\begin{equation}
\begin{aligned}
&{\mathop{\rm  min\ }\limits_{\textbf{D}[m]}}\sum\nolimits_{m=1}^{M}\lVert \textbf{P}[m]-\textbf{S}\textbf{F}[m]\textbf{D}[m]\rVert_{F}^2\\ 
&\mathrm{s.t.}\quad\textbf{D}[m]^{H}\textbf{D}[m]=\textbf{I}_{N_s}, \forall m.
\end{aligned}
\label{problem_P_D_semi}
\end{equation}
The solution to \eqref{problem_P_D_semi}, which is called the orthogonal procrustes problem, is given as~\cite{7397861,DAoSA_JSAC_2020} 
\begin{equation}
\textbf{D}[m]=\widehat{\textbf{V}}_{N_s}[m]\widehat{\textbf{U}}[m]^{H},
\label{solution_D}
\end{equation}
where $L_t\times L_t$- and $N_s\times N_s$-dimensional $\widehat{\textbf{V}}[m]$ and $\widehat{\textbf{U}}[m]$ are from the SVD of $\textbf{P}[m]^{H}\textbf{S}\textbf{F}[m]$, yielding that $\textbf{P}[m]^{H}\textbf{S}\textbf{F}[m]=\widehat{\textbf{U}}[m]\widehat{\bm \Sigma}[m]\widehat{\textbf{V}}[m]^{H}$, and $\widehat{\textbf{V}}_{N_s}[m]$ is the first $N_s$ columns of $\widehat{\textbf{V}}[m]$.

To this end, we can alternatively solve $\textbf{S}$ and $\textbf{D}[m]$ via \eqref{solution_S_i} and \eqref{solution_D} until convergence for DS-FTTD architecture. After that, we enforce the transmit power constraint to $\textbf{D}[m]$ such that $\textbf{D}[m]=\frac{\lVert\textbf{P}[m]\rVert_{F}}{\ \ \left\lVert{\textbf{S}\textbf{F}[m]\textbf{D}[m]}\right\rVert_{F}}\textbf{D}[m]$.
The pseudo code of RD algorithm is described in \textbf{Algorithm 1}. 
Since problem~\eqref{subproblem_SFD_2} can be solved with different $i$ in parallel, we use `Parallel for' in Step 04 to represent this parallel optimization procedure.

\begin{table}
	\centering
	\begin{tabular}{p{230pt}}
		\hline \textbf{Algorithm 1: RD algorithm} \\
		\hline \textbf{Input:} $L_t$, $Q$, $\textbf{P}[m]$, $m=1,2,...,M$\\
		\quad01:\quad Initialize $\textbf{S}$ randomly under the constraints in \eqref{problem_SFD} \\
		\quad02:\quad Initialize $\textbf{D}[m]$ as \eqref{solution_D}, $m=1,2,...,M$ \\
		\quad03:\quad \textbf{Repeat}\\
		\quad04:\quad \textbf{Parallel for} $i=1:{N_t}$\\
		\quad05:\quad \quad Update $\textbf{S}_i$ via \eqref{solution_S_i} \\		
		\quad06:\quad \textbf{end for}\\
		\quad07:\quad Update $\textbf{D}[m]$ through \eqref{solution_D}, $m=1,2,...,M$ \\
		\quad08:\quad \textbf{Until convergence} \\
		\quad09:\quad Normalize each $\textbf{D}[m]$ as  $\textbf{D}[m]=\frac{\lVert\textbf{P}[m]\rVert_{F}}{\ \ \left\lVert{\textbf{S}\textbf{F}[m]\textbf{D}[m]}\right\rVert_{F}}\textbf{D}[m]$\\
		\textbf{Output:} ${\textbf{S}}$ and $\textbf{D}[m]$, $m=1,2,...,M$\\
		\hline
	\end{tabular}
	\vspace{-7.5mm}
\end{table}
\subsubsection{\textbf{Convergence analysis}}
It is worth noting that we add a semi-unitary constraint to $\textbf{D}[m]$. Under this semi-unitary constraint, when $\textbf{S}$ is fixed, the solution of $\textbf{D}[m]$ in~\eqref{solution_D} is optimal to minimize the objective function~\eqref{problem_SFD_objective}. Moreover, when $\textbf{D}[m]$ is fixed, we obtain the optimal $\textbf{S}$ through~\eqref{solution_S_i}. Therefore, in \textbf{Algorithm 1}, during the iterations from Step 03 to Step 08, the objective function does not increase, i.e., the convergence to a local optimal point can be ensured. The numerical results in Sec.~\ref{section_simulation}-D show that, the RD algorithm converges fast after about 8 iterations.
\subsubsection{\textbf{Computational complexity analysis}}
We denote the number of iterations of the RD algorithm as $K$.
The computational complexity of the RD algorithm is divided into three parts, i.e.,  i) $K$ times of the calculation of $\textbf{S}$, ii) $K$ times of the calculation of $\textbf{D}[m]$, $m=1,2,...,M$, iii) one time of the normalization in Step 09.

The computation of $\textbf{S}$ involves $N_t$ times calculation of $\sum\nolimits_{m=1}^{M}(-2{\rm Re}(\textbf{F}[m]\textbf{D}[m]\textbf{P}_i[m]^H)$, once calculation of $\sum\nolimits_{m=1}^{M}{\rm diag}(\textbf{F}[m]\textbf{D}[m]\textbf{D}[m]^H\textbf{F}[m]^H)$, and $N_t$ times running of sorting algorithm, whose computational complexity can be represented as $\mathcal{O}(MN_tL_t^2QN_s)$, $\mathcal{O}(ML_t^2QN_s)$, and $\mathcal{O}(N_tL_tQ)$, respectively. In THz UM-MIMO systems, the number of antennas $N_t$ is usually much larger than the number of RF chains $L_t$ and the number of data streams $N_s$ such that $N_t$ has a significant impact on computational complexity~\cite{DAoSA_JSAC_2020}. Therefore, the computational complexity of solution of $\textbf{S}$ can be treated as $\mathcal{O}(MN_tL_t^2QN_s)$. 

The computation of $\textbf{D}[m]$ involves the calculation of $\textbf{P}[m]^{H}\textbf{S}\textbf{F}[m]$, the calculation of SVD of $\textbf{P}[m]^{H}\textbf{S}\textbf{F}[m]$, and the calculation of $\widehat{\textbf{V}}_{N_s}[m]\widehat{\textbf{U}}[m]^{H}$, whose computational complexity is $\mathcal{O}(N_tL_t^2Q)$, $\mathcal{O}(L_t^3)$, and $\mathcal{O}(L_tN_s^2)$, respectively. Since $N_t$ is much larger than $L_t$ and $N_s$, the computational complexity of the solution $\textbf{D}[m]$ for $m=1,2,...,M$ can be treated as $\mathcal{O}(MN_tL_t^2Q)$. Furthermore, the computational complexity of the normalization in step 09 is $\mathcal{O}(N_tL_t^2Q)$.

Consequently, by combining the above terms, the total computational complexity of the RD algorithm is $\mathcal{O}(KMN_tL_t^2QN_s)$, which grows linearly with the number of antennas $N_t$ and is low.
\section{Simulation Results and Analysis}
\label{section_simulation}
In this section, we comprehensively evaluate the performance of the proposed DS-FTTD architecture with the RD algorithm in the THz wideband UM-MIMO systems. The simulation setup is given in Sec.~\ref{section_simulation}-A. We first evaluate the performance of the DS-FTTD architecture versus the number of FTTDs in Sec.~\ref{section_simulation}-B, including the array gain, spectral efficiency, and energy efficiency. Then, we analyze the spectral efficiency and energy efficiency of the DS-FTTD architecture with varying transmit power and number of antennas, in Sec.~\ref{section_simulation}-C. Moreover, in Sec.~\ref{section_simulation}-D, we elaborate the convergence of the proposed RD algorithm and analyze the impact of the imperfect CSI on spectral efficiency of the DS-FTTD architecture with RD algorithm.

\subsection{Simulation Setup}
The simulation setup is presented as follows, unless stated otherwise. The central frequency is $f_c=300$ GHz, where the bandwidth is $B=50$ GHz. The number of carriers is $M=50$. The communication distance is 50m. The number of multipath is less than 5 and the generation of the THz multipath as well as the path gain and directions follow the ray-tracing method in~\cite{DAoSA_JSAC_2020}. We consider UPA at both the transmitter and receiver. The number of antennas at transmitter and receiver is $N_t=N_r=1024$. The number of RF chains and data streams is $L_t=N_s=4$.
We consider to set the antenna beamwidth in (3) as $\frac{2\pi}{3}$ and $\frac{\pi}{4}$ at the azimuth direction and elevation direction to support a wide coverage region, which is usually used at the base station~\cite{7737056}. Together with the ultra-sharp array beam pattern generated by beamforming, the resulting overall beam pattern has high gain and is highly directional and adjustable within the coverage region.
The antenna gain is $G_0=8.8$ dBi calculated by~\eqref{antenna_gain_model}.

\subsubsection{\textbf{Competitors of proposed DS-FTTD architecture}}
We compare our proposed scheme, i.e., the DS-FTTD architecture with the proposed RD algorithm, with the following schemes, where `FC-PS', `DS-PS', and `AoSA-PS' represent the typical FC, DS, and AoSA architectures with phase shifters in~\cite{9398864,9110865,7880698}.
`GoSA' refers to the group of subarrays (GoSA) architecture in~\cite{9557817}. Multiple overlapping/non-overlapping GoSA architectures have been proposed in~\cite{9557817} for the trade-offs between the spectral efficiency and complexity. In our simulations, we consider a low-complexity partially-connected GoSA architecture, in which the antennas connected with each RF chain are divided into multiple non-overlapping subarrays. Each subarray contains $Q_{GoSA}$ antennas which are connected with the same phase shifter to reduce the hardware complexity and power consumption. We set $Q_{GoSA}=4$ in simulations.

\textbf{i)} FC-TTD architecture with modified wideband PE-AltMin algorithm in~\cite{7397861}. Note that the original PE-AltMin algorithm is designed for FC-PS architecture. By changing the optimization of phase shift to time delay, it can be modified to work for FC-TTD architecture. 

\textbf{ii)} TTD-aided architecture with wideband delay-phase precoding algorithm in~\cite{daiDPP}. 

\textbf{iii)} FC-PS architecture with wideband virtual subarray algorithm in~\cite{9398864}.

\textbf{iv)} DS-PS architecture with wideband beamforming algorithm in~\cite{9110865}. 

\textbf{v)} AoSA-PS architecture with wideband beamforming algorithm in~\cite{7880698}. 

\textbf{vi)} GoSA architecture with beam split correction algorithm in~\cite{9557817}.
\subsubsection{\textbf{Power consumption of different architectures}}
\begin{table}
	\centering
	\captionsetup{font={footnotesize}}
	\caption{Power consumption of DS-FTTD and the existing architectures at Tx.}
	\begin{tabular}{cc} 
		\hline
		Architecture&Power consumption\\
		\hline Proposed DS-FTTD&${\rm P_{u}}+{\rm P_{FTTD}}\rm N^{a}_{FTTD}+{\rm P_{SW}}N_t+{\rm P_{PD}}{\rm N_{PD}}$\\
		FC-TTD&${\rm P_{u}}+{\rm P_{TTD}}N_{t}L_t+{\rm P_{PD}}L_t+{\rm P_{PC}}N_t$\\
		TTD-aided& ${\rm P_{u}}+{\rm P_{TTD}}N_k+{\rm P_{PS}}N_{t}L_t+{\rm P_{PD}}(L_t+N_k)+{\rm P_{PC}}N_t$\\
		FC-PS&${\rm P_{u}}+{\rm P_{PS}}N_{t}L_t+{\rm P_{PD}}L_t+{\rm P_{PC}}N_t$\\
		DS-PS&${\rm P_{u}}+{\rm P_{PS}}N_{t}+{\rm P_{SW}}N_t+{\rm P_{PD}}L_t$\\
		AoSA-PS&${\rm P_{u}}+{\rm P_{PS}}N_{t}+{\rm P_{PD}}L_t$\\
		GoSA&${\rm P_{u}}+{\rm P_{PS}}N_t/Q_{GoSA}+{\rm P_{PD}}(L_t+N_t/Q_{GoSA})$\\					
		\hline	
	\end{tabular}
	\label{table_power_existing_architectures}
	\vspace{-6.5mm}
\end{table}
We list the power consumption of the DS-FTTD architecture and the above existing architectures in TABLE~\ref{table_power_existing_architectures}, where the parameters are presented as follows. ${\rm P_u}={\rm P_{PA}}N_t+{\rm P_{RF}}L_t+{\rm P_{DAC}}L_t+{\rm P_{BB}}+\rho$ is the power consumption of the common part of these architectures, where $\rho$ is the transmit power. Specifically, the power consumption in the unit of mW around $300$ GHz of power amplifier, RF chain, DAC, baseband, phase shifter, adjustable TTD, FTTD, switch, power divider, and power combiner is ${\rm P_{PA}}=60$, ${\rm P_{RF}}=26$, ${\rm P_{DAC}}=110$, ${\rm P_{BB}}=200$, ${\rm P_{PS}}=42$, ${\rm P_{TTD}}=80$, ${\rm P_{FTTD}}=30$, ${\rm P_{SW}}=10$, ${\rm P_{PD}}=6.6$, and ${\rm P_{PC}}=6.6$, respectively~\cite{DAoSA_JSAC_2020,Tan2020PrecodingSurvey,6005342,7436794,8733134}. The coefficients of the power consumption of these devices in TABLE~\ref{table_power_existing_architectures} denote the quantities of these devices, where ${\rm N_{PD}}=L_t+\rm N^{a}_{FTTD}$. 
The number of active FTTDs $\rm N^{a}_{FTTD}$ is equal to the quantity of the non-zero columns of $\textbf{S}$.
$N_k$ for TTD-aided architecture is set as 128 in the simulations.
\subsection{Performance of the DS-FTTD Architecture versus the Number of FTTDs}
\begin{figure}
	\setlength{\belowcaptionskip}{0pt}
	\centering
	\captionsetup{font={footnotesize}}
	\subfigure[Array gain with different frequencies versus $Q$, $\phi_0=45^{\circ}$, $\theta_0=30^{\circ}$.]{
		\includegraphics[scale=0.286]{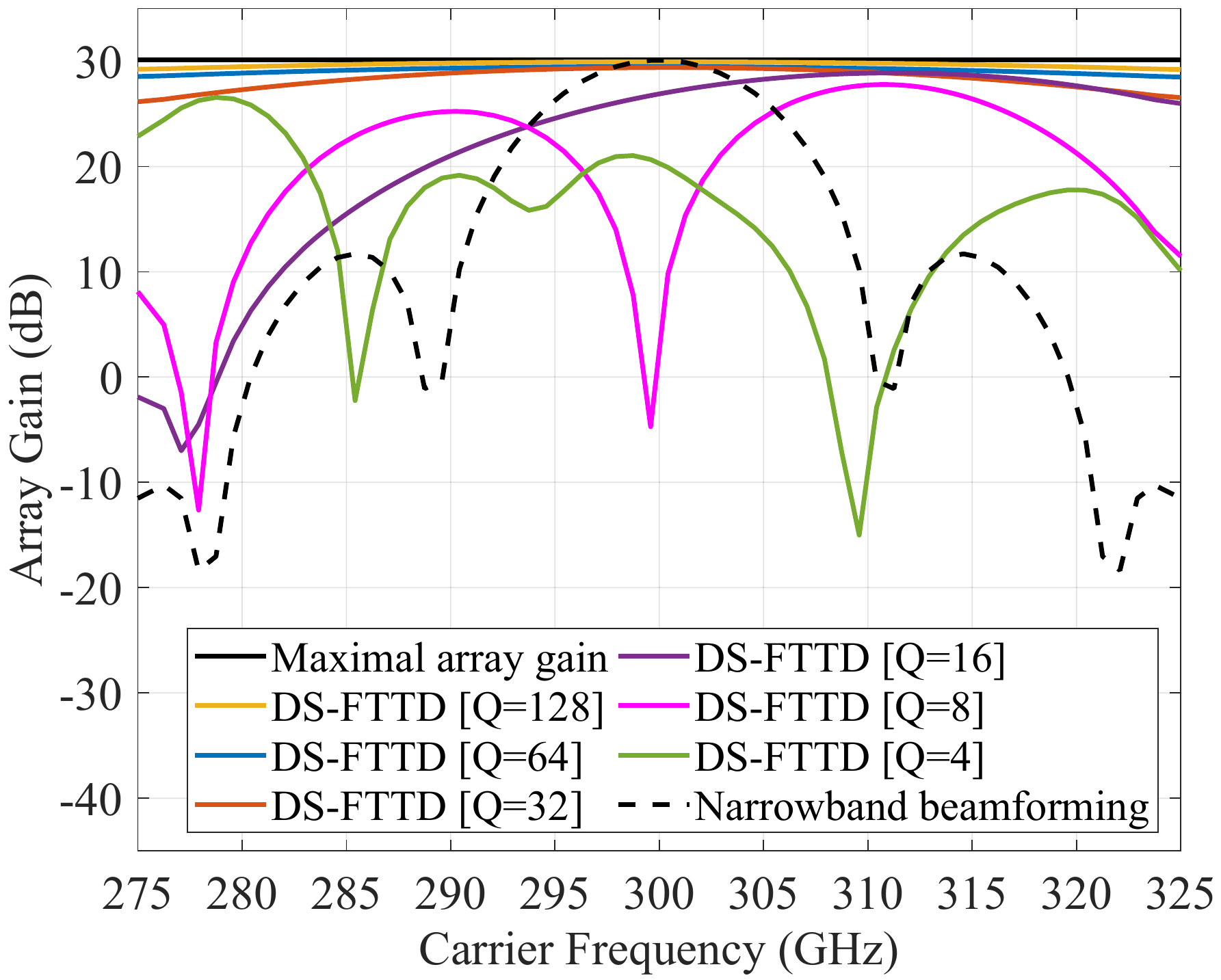}}	
	\subfigure[Spectral efficiency versus $Q$.]{
		\includegraphics[scale=0.282]{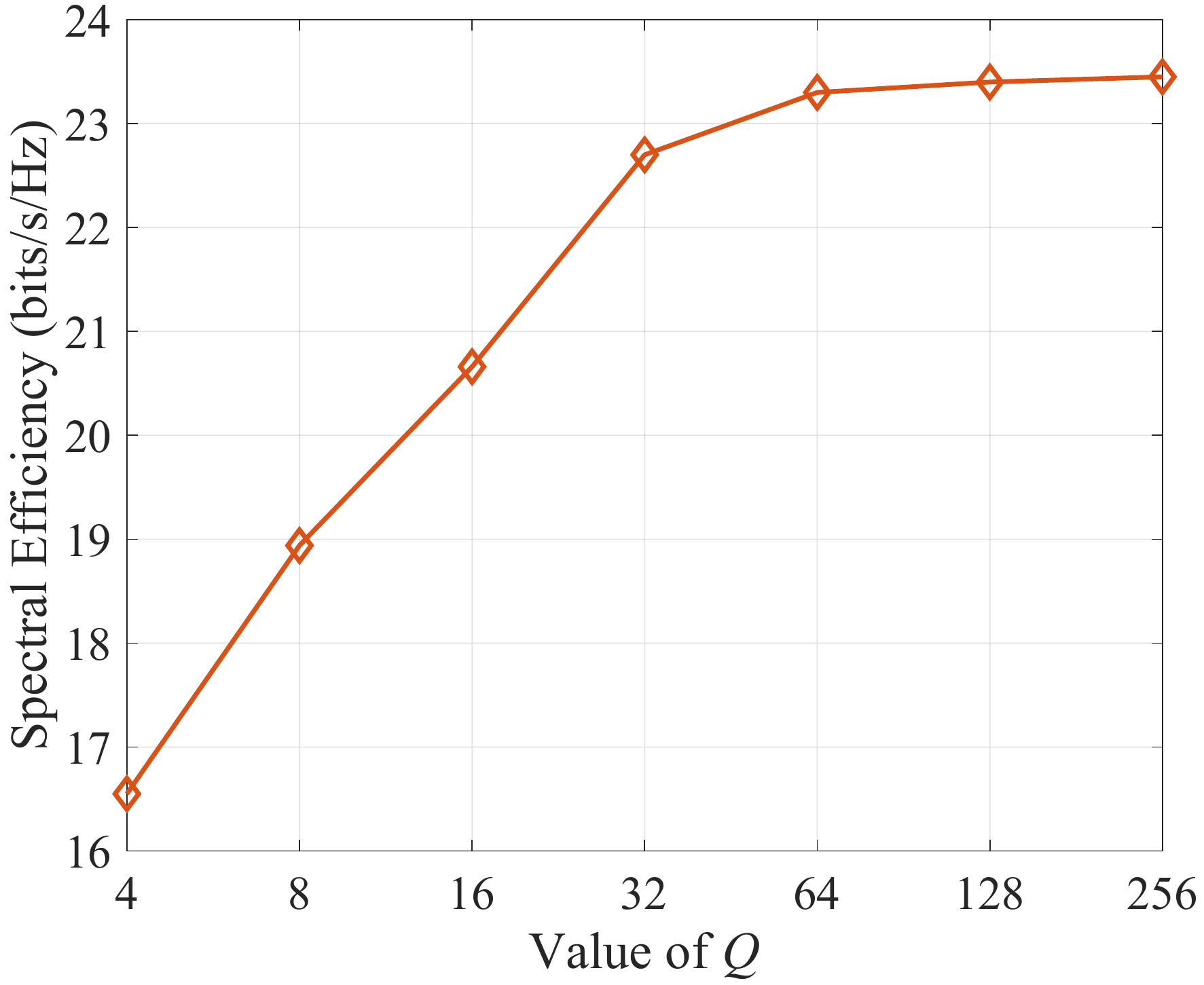}}
	\subfigure[Energy efficiency versus $Q$.]{
		\includegraphics[scale=0.282]{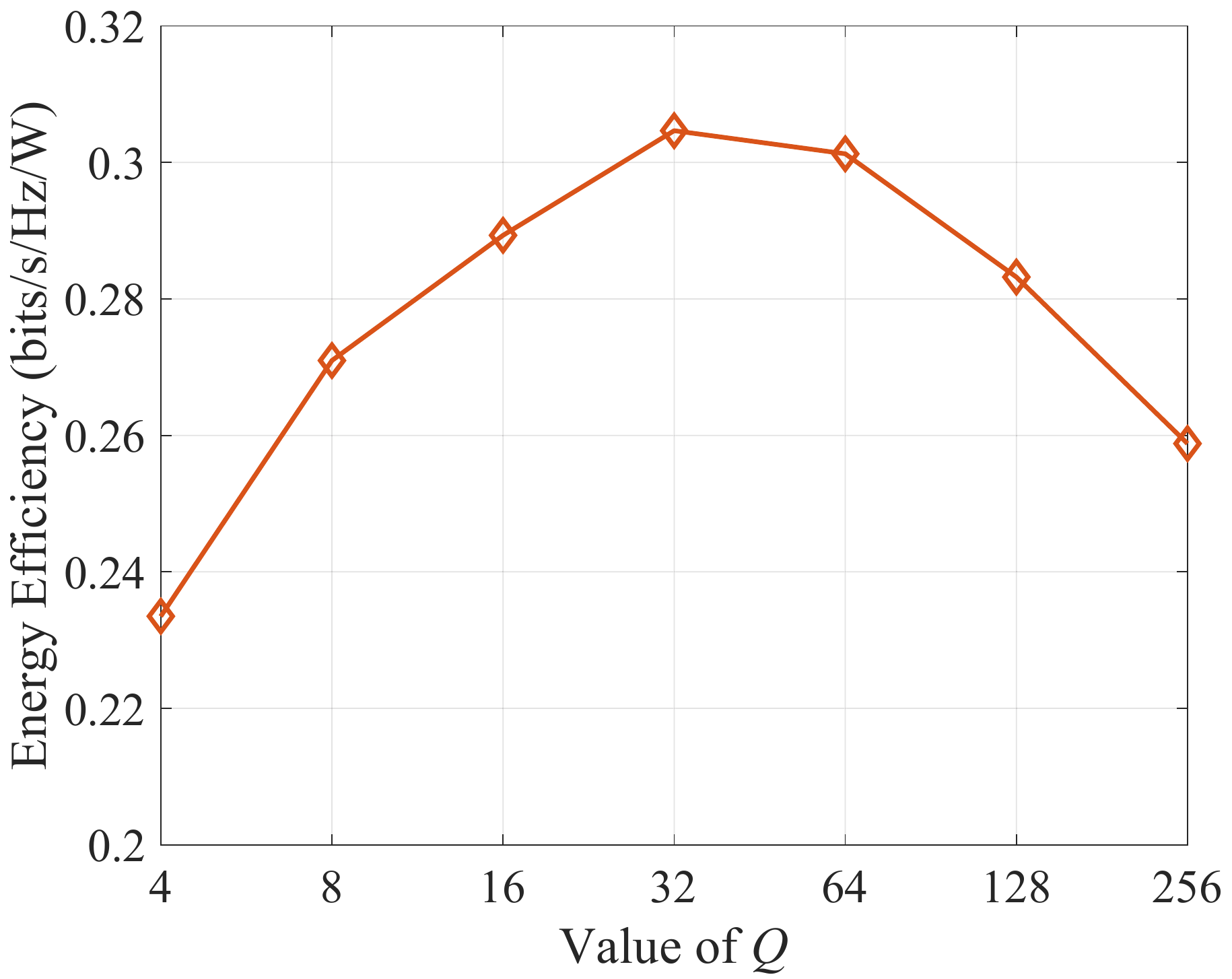}}
	\caption{The array gain at transmitter, spectral efficiency, and energy efficiency of the proposed DS-FTTD architecture versus $Q$. $N_t=N_r=1024$. The transmit power is $\rho=20$ dBm.}
	\label{fig_AF_SE_EE_Q}
	\vspace{-7.5mm}
\end{figure}
First, we aim to evaluate the performance of the DS-FTTD architecture versus the number of FTTDs, including the array gain, spectral efficiency, and energy efficiency. The number of FTTDs in DS-FTTD architecture is $L_tQ$ such that we vary the value of $Q$. 
Fig.~\ref{fig_AF_SE_EE_Q}(a) shows the array gain of the DS-FTTD architecture from 275 GHz to 325 GHz. The ideal maximal array gain of an array with $N_t=1024$ antennas is $10{\rm log}_{10}N_t=30.1$ dB over all frequencies.
By using the narrowband central frequency steering vector in~\eqref{steering_UPA_PS_1}, the phase shifters-based narrowband beamforming can only achieve the maximal array gain at the central frequency. While for other frequencies, the array gain is substantially reduced, e.g., 40 dB loss at 275 GHz. The average array gain of narrowband beamforming is only 9.8 dB, which is more than 20 dB lower than the maximal array gain. In the proposed DS-FTTD architecture with various $Q$, the average array gain is always higher than the narrowband beamforming. Moreover, with larger $Q$, more candidate time delays can be provided to each antenna for selection such that the array gain increases. Specifically, the average array gain over all frequencies of the DS-FTTD architecture from $Q=4$ to $Q=128$ is 12.6 dB, 16.8 dB, 21.3 dB, 27.9 dB, 29.0 dB, and 29.7 dB, respectively. When $Q\geq 64$, the average array gain loss compared to the maximal array gain is about 1 dB, which is very small. Therefore, the DS-FTTD architecture can address the beam squint efficiently.

As shown in Fig.~\ref{fig_AF_SE_EE_Q}(b), we evaluate the spectral efficiency of the DS-FTTD architecture versus $Q$.  
With larger $Q$, i.e., more FTTDs, the spectral efficiency of the DS-FTTD architecture increases. Particularly, when $Q$ exceeds 128, the spectral efficiency hardly enhances with larger $Q$. Therefore, $Q=128$ is a proper value for DS-FTTD architecture to achieve a high spectral efficiency. 
As shown in Fig.~\ref{fig_AF_SE_EE_Q}(c), the energy efficiency has a different trend from the array gain and the spectral efficiency. 
With larger $Q$, i.e., more FTTDs, the array gain and spectral efficiency increase. By contrast, the power consumption of FTTDs also increases with larger $Q$. When $Q<32$, with more FTTDs, the spectral efficiency increases faster than the improvement of power consumption such that the energy efficiency grows. When $Q>32$, the increasing of power consumption dominates the trend of the energy efficiency such that the energy efficiency reduces. As a result, the highest energy efficiency of DS-FTTD is achieved at $Q=32$. To achieve high energy efficiency, we should not use too many FTTDs and $Q=32$ is a proper value, which will be used in the following simulations.

\subsection{Spectral Efficiency and Energy Efficiency of the Proposed DS-FTTD Architecture}
\begin{figure}
	\setlength{\belowcaptionskip}{0pt}
	\centering
	\captionsetup{font={footnotesize}}
	\subfigure[Spectral efficiency versus transmit power $\rho$.]{
		\includegraphics[scale=0.4]{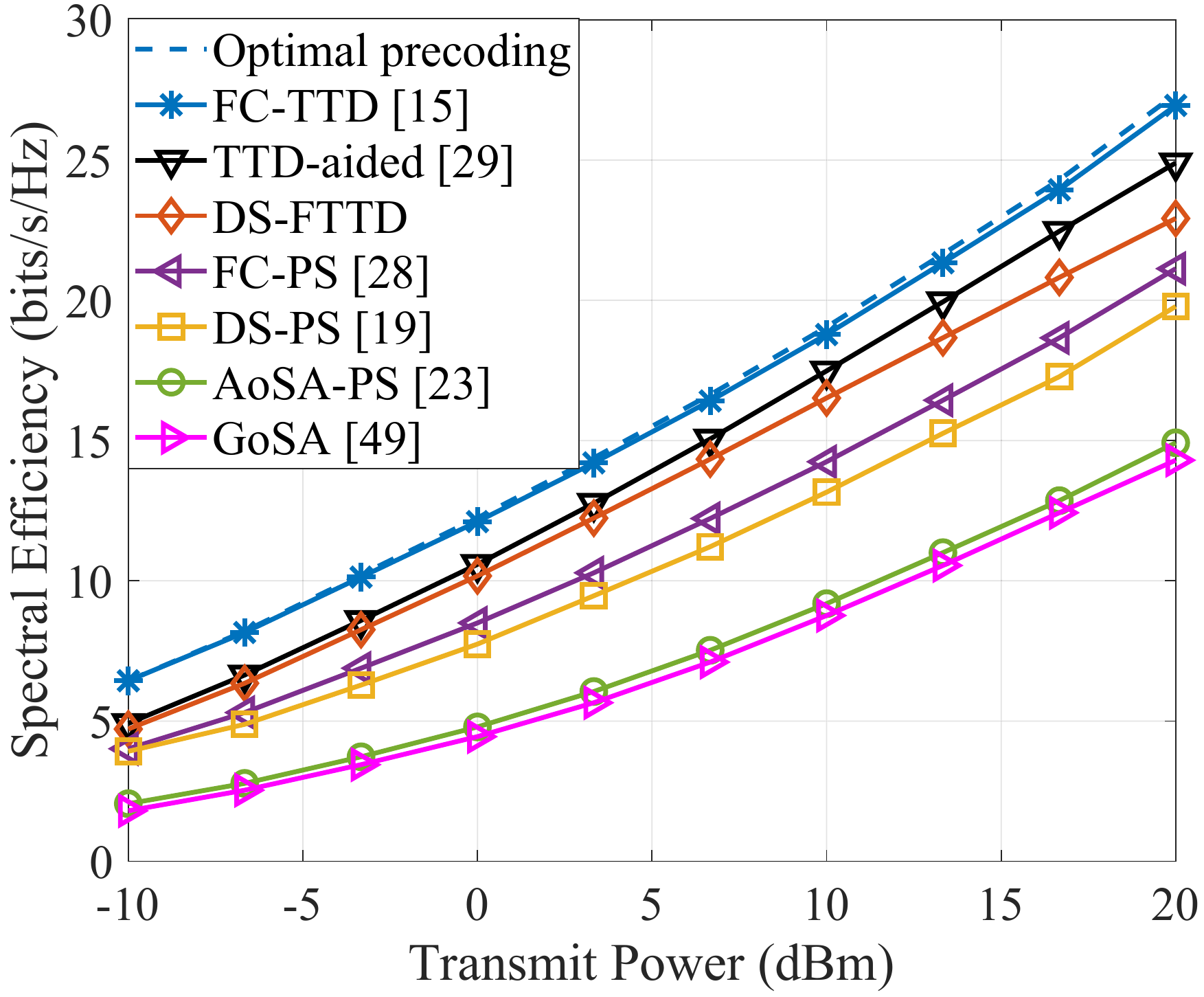}}
	\subfigure[Energy efficiency versus spectral efficiency, $\rho=20$ dBm.]{
		\includegraphics[scale=0.4]{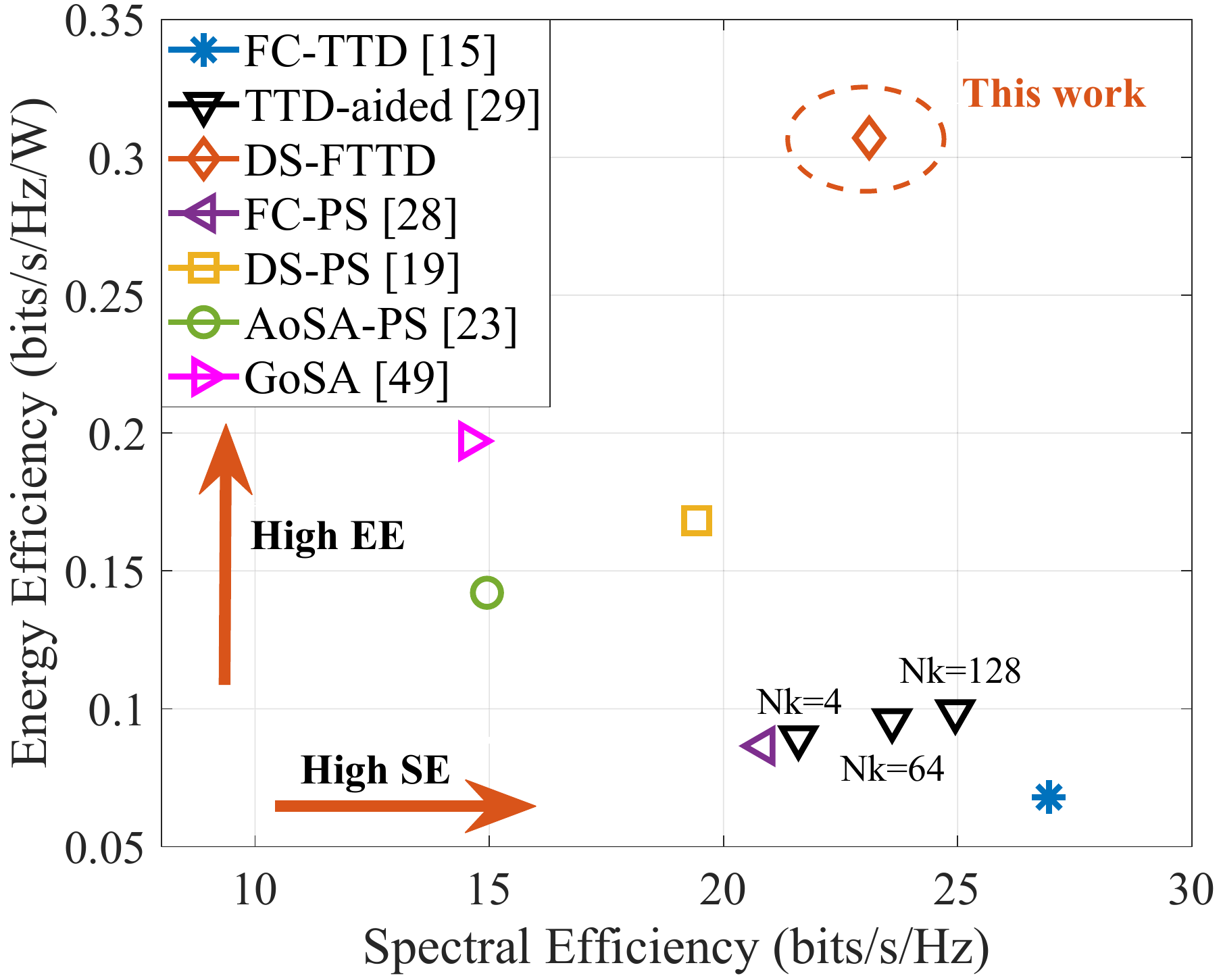}}
	\caption{The spectral efficiency and energy efficiency of the DS-FTTD architecture, $N_t=N_r=1024$, $Q=32$.}
	\label{fig_SE_EE_rho}
	\vspace{-8.5mm}
\end{figure}

\begin{figure}
	\setlength{\belowcaptionskip}{0pt}
	\centering
	\captionsetup{font={footnotesize}}
	\subfigure[Spectral efficiency versus number of antennas.]{
		\includegraphics[scale=0.4]{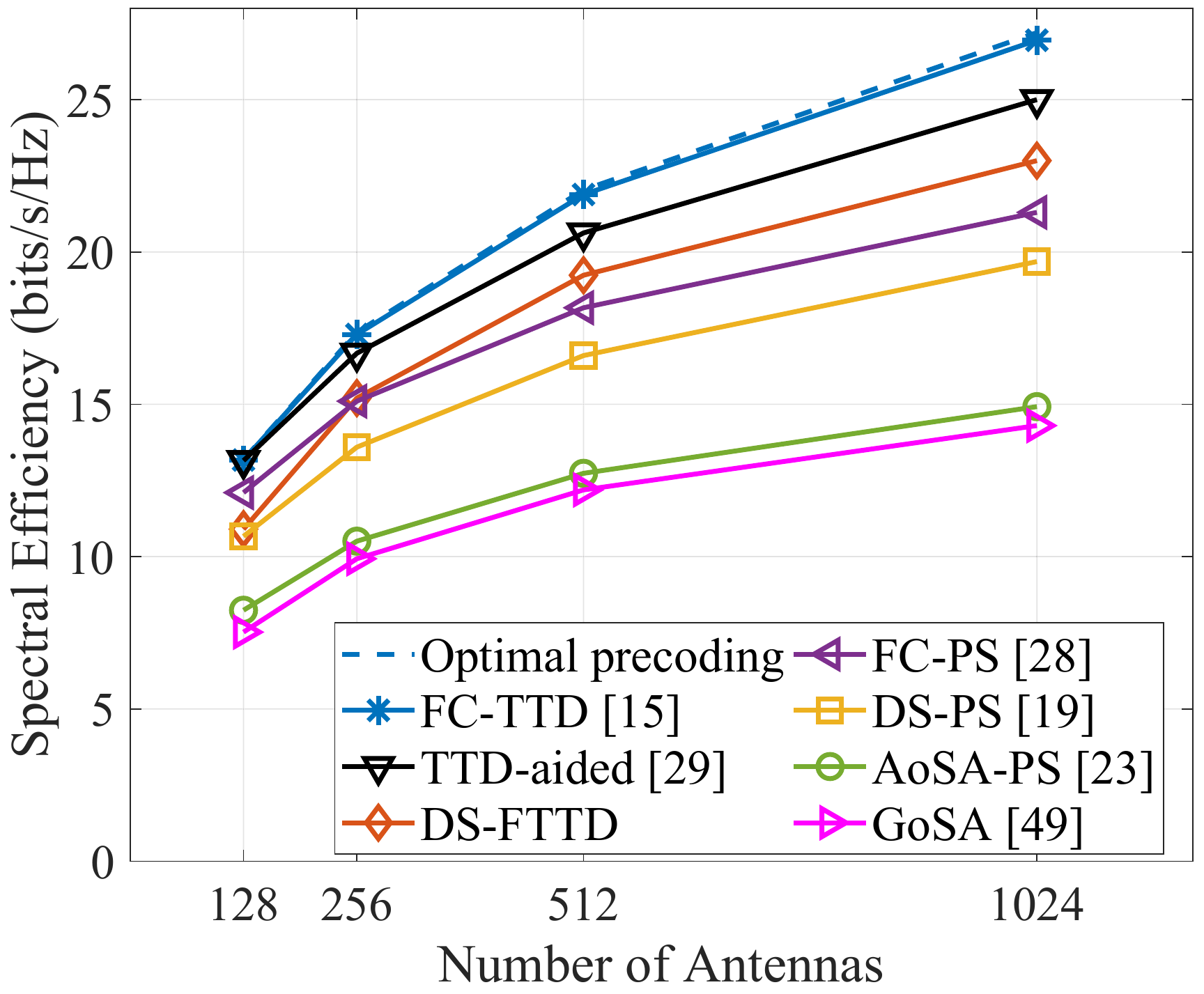}}
	\subfigure[Energy efficiency versus number of antennas.]{
		\includegraphics[scale=0.4]{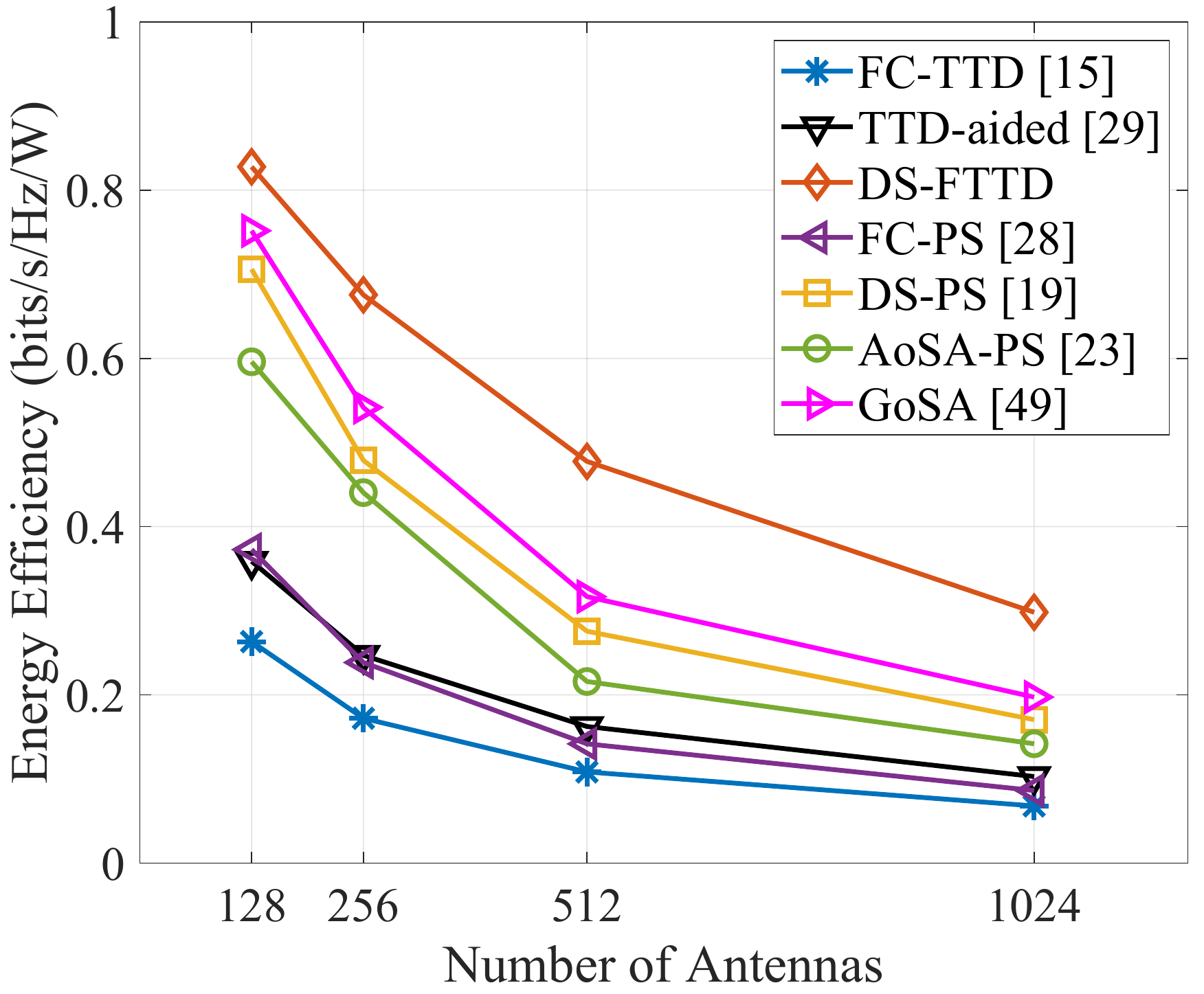}}
	\caption{The spectral efficiency and energy efficiency versus number of antennas of the DS-FTTD architecture, $N_t=N_r$, $Q=32$, $\rho=20$ dBm.}
	\label{fig_SE_EE_Nt}
	\vspace{-8.5mm}
\end{figure}

We compare the spectral efficiency and energy efficiency of the DS-FTTD architecture with the existing architectures in Fig.~\ref{fig_SE_EE_rho} and Fig.~\ref{fig_SE_EE_Nt}.
The energy efficiency is defined as the ratio between the spectral efficiency and the power consumption in TABLE~\ref{table_power_existing_architectures}. 

Fig.~\ref{fig_SE_EE_rho}(a) and Fig.~\ref{fig_SE_EE_rho}(b) evaluate the spectral efficiency and energy efficiency of the DS-FTTD architecture using the proposed RD algorithm.
As shown in Fig.~\ref{fig_SE_EE_rho}(a), the optimal precoding refers to the optimal precoding matrix $\textbf{P}[m]$ given in~\eqref{problem_SFD_objective}, whose spectral efficiency is the upper bound of all hybrid beamforming architectures. The spectral efficiency of the FC-TTD architecture approaches that of the optimal precoding. Compared to the optimal precoding and the FC-TTD architecture, the spectral efficiency of DS-FTTD with the proposed RD algorithm is about 4~bits/s/Hz lower when $\rho=20$ dBm.
Compared to the TTD-aided architecture, the spectral efficiency of DS-FTTD is lower by 2 bits/s/Hz.
The spectral efficiency of the DS-FTTD architecture is higher than the FC-PS and DS-PS architectures by 2 bits/s/Hz and 3 bits/s/Hz when $\rho=20$ dBm, which reveals that the proposed DS-FTTD architecture has better capability to combat the beam squint effect.
The~AoSA and GoSA are two subarray-based architectures with low hardware complexity and power consumption. Particularly, owing to the partial-connection and the share of the same phase shifter of the $Q_{GoSA}$ antennas in each subarray, the GoSA architecture only uses $\frac{N_t}{Q_{GoSA}}$ phase shifters, which are much less than that of the existing phase shifters-based architectures. As shown in Fig.~5(a), with low hardware complexity, the spectral efficiencies of the AoSA architecture and GoSA architecture are low, e.g., 8 bits/s/Hz and 8.3 bits/s/Hz lower than the proposed DS-FTTD architecture when $\rho=20$ dBm.

Although the spectral efficiency of the DS-FTTD architecture is lower than the FC-TTD and TTD-aided counterparts with $N_k=128$, the energy efficiency of DS-FTTD is much higher than these two architectures, owing to the low-cost FTTDs and switches, as shown in Fig.~\ref{fig_SE_EE_rho}(b). 
Moreover, we evaluate the performance of the TTD-aided architecture with different numbers of TTDs, $N_k$. Specifically, when $N_k=64$ and $N_k=4$, the spectral efficiency is lower than that with $N_k=128$, due to the less TTDs to adjust time delay. Compared to our DS-FTTD architecture, the TTD-aided with $N_k=128$, $N_k=64$, and $N_k=4$ achieves 2 bits/s/Hz higher, 0.6 bits/s/Hz higher, and 1.5 bits/s/Hz lower spectral efficiency, respectively. 
Note that the power consumption of TTD-aided architecture is dominated by the phase shifters, whose quantity is very large, i.e., $N_tL_t=4096$~\cite{9398864,daiDPP}. Therefore, the power consumption of the TTD-aided architecture is reduced by only $2.2\%$ and $4.3\%$, by reducing $N_k$ from $128$ to $64$ and $4$. As a combined effect of the spectral efficiency and power consumption, the energy efficiency of the TTD-aided with $N_k=64$ and $N_k=4$ is $3.3\%$ and $9.6\%$ lower than that of $N_k=128$. Remarkably, the energy efficiency of our proposed DS-FTTD architecture is $243\%$, $221\%$, and $210\%$ higher than the TTD-aided architecture with $N_k=4$, $64$, and $128$, respectively.
Compared to the DS-PS, FC-PS, and AoSA-PS architectures, the DS-FTTD architecture achieves both improved energy efficiency and spectral efficiency. 
Although the GoSA architecture has lower power consumption than the proposed DS-FTTD architecture, the DS-FTTD architecture achieves a much higher spectral efficiency, which results in 0.1 bits/s/Hz/W higher energy efficiency, as shown in Fig.~\ref{fig_SE_EE_rho}(b).
Consequently, compared to the existing architectures, the DS-FTTD architecture with the proposed RD algorithm can achieve significantly higher energy efficiency and good spectral efficiency.

Fig.~\ref{fig_SE_EE_Nt}(a) and Fig.~\ref{fig_SE_EE_Nt}(b) evaluate the spectral efficiency and energy efficiency versus the number of antennas. As shown in Fig.~\ref{fig_SE_EE_Nt}(a), by increasing the number of antennas, the spectral efficiencies of all architectures increase. 
With a large number of antennas, e.g., $N_t\geq256$, the spectral efficiency of the DS-FTTD architecture is always higher than the FC-PS, DS-PS, AoSA-PS, and GoSA architectures, while is lower than the optimal precoding, FC-TTD, and TTD-aided architectures.
As shown in Fig.~\ref{fig_SE_EE_Nt}(b), the energy efficiency of DS-FTTD architecture is higher than the other architectures with various numbers of antennas. Compared to other architectures, the superiority of energy efficiency of the DS-FTTD grows with the number of antennas. For instance, with 128 antennas, the energy efficiency of DS-FTTD is 7\%, 15\%, and 36\% higher than the GoSA, DS-PS, and AoSA-PS architectures, respectively. By contrast, with 1024 antennas, the improvement of energy efficiency of DS-FTTD enlarges to 55\%, 75\%, and 110\%, respectively.
In THz UM-MIMO systems, the number of antennas is usually large, e.g., $N_t>512$. Hence, the proposed DS-FTTD is suitable and energy-efficient for THz UM-MIMO systems.
The spectral efficiency of the DS-FTTD architecture is 4 bits/s/Hz and 2 bits/s/Hz lower than FC-TTD and TTD-aided architectures with 1024 antennas, respectively. However, considering the lower hardware complexity and much higher energy efficiency, the spectral efficiency loss is acceptable.

\begin{figure}
	\setlength{\belowcaptionskip}{0pt}
	\centering
	\includegraphics[scale=0.4]{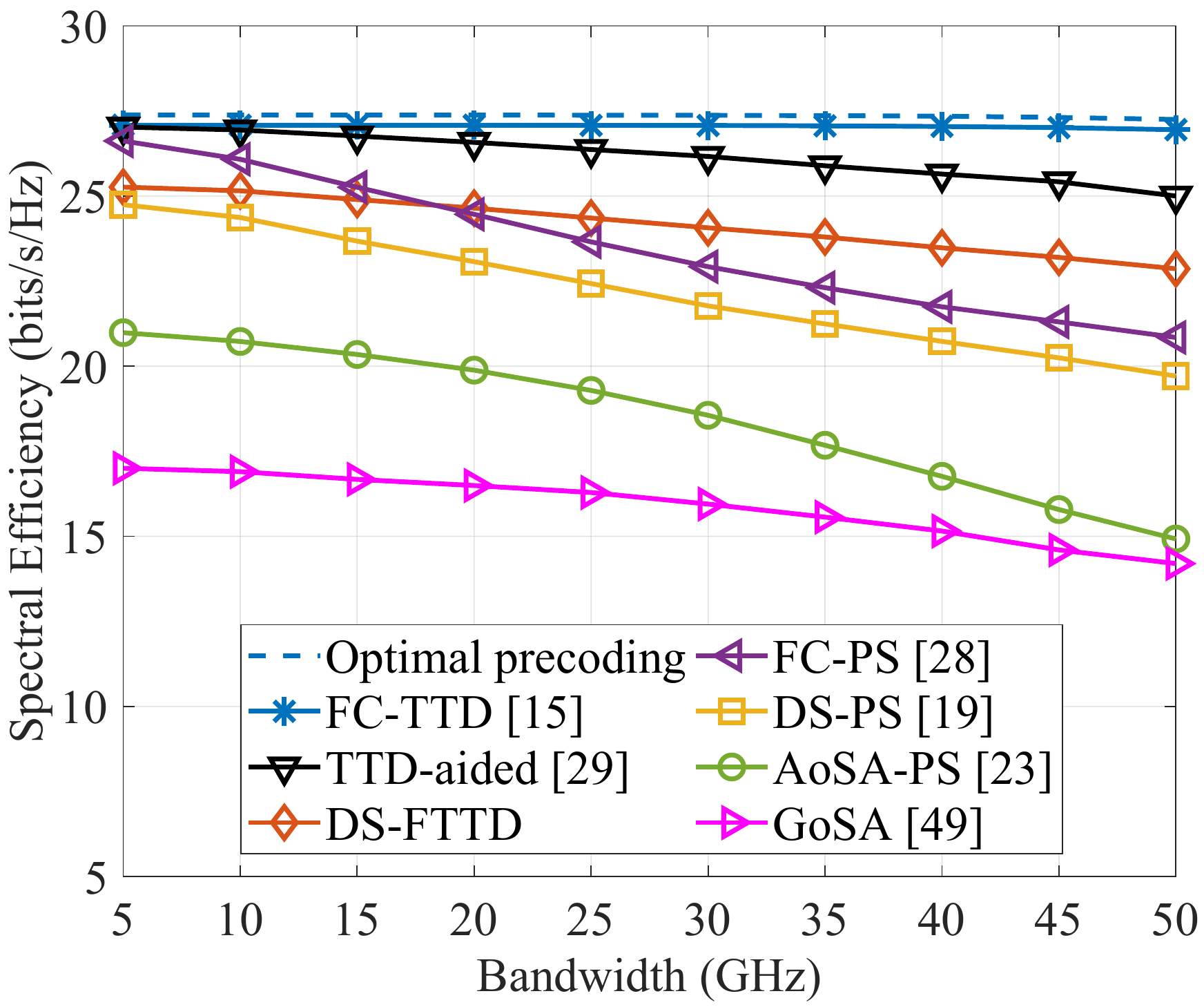} 
	\captionsetup{font={footnotesize}}
	\caption{The spectral efficiency of the DS-FTTD architecture versus the bandwidth $B$. $N_t=N_r=1024$, $Q=32$. The transmit power per GHz is set as a constant $\frac{\rho}{B}=\frac{0.1 \ \rm W}{50 \ \rm GHz}$ to ensure a constant SNR when varying the bandwidth.}  
	\label{fig_SE_B} 
	\vspace{-8.5mm}
\end{figure}

Fig.~\ref{fig_SE_B} evaluates the spectral efficiency of different architectures by varying the bandwidth. With larger bandwidth, the beam squint becomes severer and the spectral efficiencies of most architectures reduce.
Specifically, the FC-TTD architecture can address the beam squint well and achieve similar spectral efficiency with the optimal precoding matrix $\textbf{P}[m]$ in~\eqref{problem_SFD_objective}. The spectral efficiencies of the DS-FTTD architecture and the TTD-aided architecture reduce slowly with the bandwidth, which reveal that both FTTDs in the DS-FTTD architecture and TTDs in the TTD-aided architecture can reduce the impact of the beam squint problem effectively. For FC-PS, DS-PS, and AoSA-PS architectures, the spectral efficiencies reduce rapidly with the bandwidth. Furthermore, the spectral efficiency of the GoSA architecture decreases slowly with the bandwidth, which demonstrates that the beam split correction algorithm proposed in~\cite{9557817} is effective to combat the beam squint problem.

\subsection{Convergence of RD Algorithm and Impact of Imperfect CSI of the Proposed DS-FTTD Architecture}
\begin{figure}
	\centering
	\begin{tabular}{cc}
		\begin{minipage}[t]{0.46\linewidth}
			\includegraphics[width = 1\linewidth]{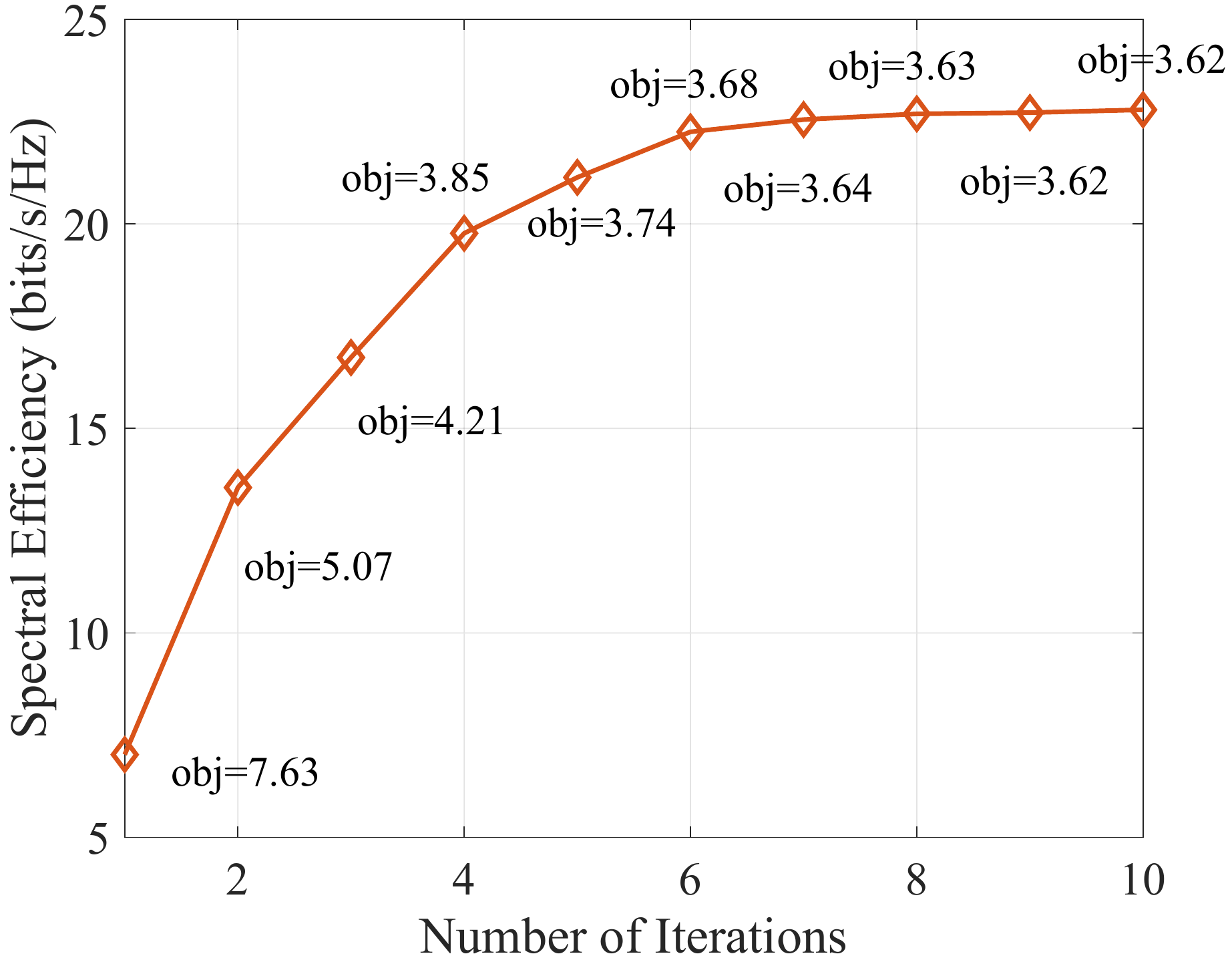}
			\captionsetup{font={footnotesize}}
			\caption{Spectral efficiency versus number of iterations of the RD algorithm in DS-FTTD, $N_t=N_r=1024$, $\rho=$ 20 dBm, $Q=32$.}
			\label{fig_SE_vs_iterations}
		\end{minipage}
		\begin{minipage}[t]{0.435\linewidth}
			\includegraphics[width = 1\linewidth]{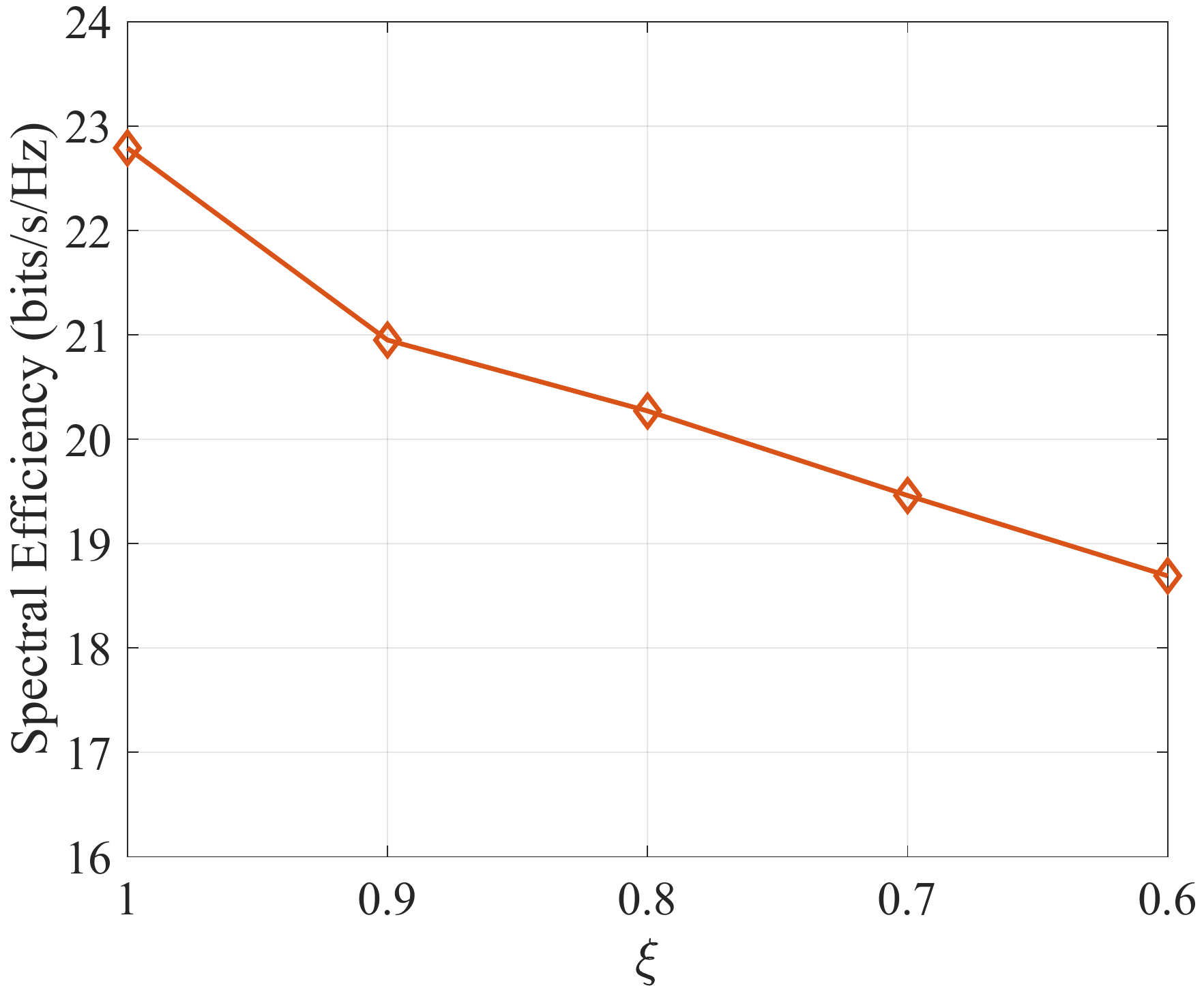}
			\captionsetup{font={footnotesize}}
			\caption{Spectral efficiency versus $\xi$ in DS-FTTD, $N_t=N_r=1024$, $\rho=$ 20 dBm, $Q=32$.}
			\label{fig_SE_vs_CSI}
		\end{minipage}
	\end{tabular}
	\vspace{-6.5mm}
\end{figure}

Fig.~\ref{fig_SE_vs_iterations} shows the spectral efficiency versus the number of iterations of the RD algorithm. 
To recall, we transform the original spectral efficiency maximization problem in~\eqref{SE_formulation} into the problem~\eqref{problem_SFD}, which is solved by the RD algorithm.
Here, `obj' denotes the average value of the objective function~\eqref{problem_SFD_objective} over all carriers. 
With more iterations, the objective function reduces and the spectral efficiency grows.
After about 8 iterations of the RD algorithm, the objective function~\eqref{problem_SFD_objective} converges.
Meanwhile, the spectral efficiency of the DS-FTTD architecture converges, which demonstrates that the transformation of the original spectral efficiency maximization problem \eqref{SE_formulation} to the Euclidean distance minimization problem \eqref{problem_SFD} is effective.

We further assess the impact of imperfect CSI on spectral efficiency in DS-FTTD architecture in Fig.~\ref{fig_SE_vs_CSI}, since the perfect CSI is intractable to obtain in practical communications. We use $\xi$ to represent the accuracy of the CSI. The imperfect CSI, i.e., the inaccurate channel matrix $\widehat{\textbf{H}}[m]$, can be represented as
\begin{equation}
	\widehat{\textbf{H}}[m]=\xi\textbf{H}[m]+\hat{e}_m\sqrt{1-\xi^2}\textbf{E}[m], \forall m,
\end{equation}
where $\textbf{H}[m]$ is the true channel matrix and $\textbf{E}[m]$ denotes the error matrix whose elements follow the i.i.d complex Gaussian distribution that $\mathcal{CN}(0,1)$. $\hat{e}_m$ is a normalization factor to ensure $\lVert\hat{e}_m\textbf{E}[m]\rVert_{F}^2=\lVert\textbf{H}[m]\rVert_{F}^2$. As shown in Fig.~\ref{fig_SE_vs_CSI}, with larger $\xi$, the spectral efficiency reduces since the error of the known channel matrix $\widehat{\textbf{H}}[m]$ becomes larger. Even with a large error, i.e., $\xi=0.6$, the DS-FTTD architecture can still achieve 82\% spectral efficiency of the DS-FTTD architecture with perfect CSI that $\xi=1$. Consequently, the proposed DS-FTTD architecture with RD algorithm is robust to imperfect CSI.

\section{Conclusion}
\label{section_conclusion}
In this work, we have proposed an energy-efficient DS-FTTD architecture to solve the beam squint problem for THz wideband hybrid beamforming.
We first investigate the channel model for THz wideband UM-MIMO systems and then analyze the severe beam squint caused by the THz peculiarities of ultra-wide fractional bandwidth and ultra-large-scale antennas array. Then, we comprehensively analyze that the power consumption, hardware complexity, and insertion loss of the proposed DS-FTTD architecture are lower than the existing FC-TTD and TTD-aided architectures. Furthermore, we propose a low-complexity RD algorithm to solve the hybrid beamforming problem for DS-FTTD architecture. The key idea is transforming the non-convex and intractable hybrid integer programming design problem to a simple ranking problem, which can be efficiently solved.

Extensive simulation results are presented to evaluate the performance of the proposed DS-FTTD architecture and the RD algorithm. By using the RD algorithm, the DS-FTTD architecture achieves the near-optimal array gain over all carriers. We analyze the energy efficiency of the DS-FTTD architecture for varying transmit power and number of antennas and show that the energy efficiency of the DS-FTTD architecture is significantly higher than the existing architectures. Furthermore, the DS-FTTD architecture with the RD algorithm is robust when the CSI is imperfect with errors. 
The DS-FTTD architecture and hybrid beamforming algorithm proposed in this work can be extended to the multi-user systems. In particular, the formulation of the multi-user sum rate and beam squint problem need to be carefully investigated in our future work.
	
\bibliographystyle{IEEEtran}
\bibliography{IEEEabrv,reference}
\end{document}